\documentclass[prr,aps,superscriptaddress,twocolumn, 10pt]{revtex4-2}

\usepackage{float}
\usepackage{graphicx}
\usepackage{dcolumn}
\usepackage{bm}
\usepackage{xcolor}
\usepackage{appendix}
\usepackage{caption}
\usepackage{subcaption}
\usepackage{multirow}
\usepackage{booktabs}
\usepackage{amsmath}
\usepackage{amsfonts}
\usepackage[normalem]{ulem}

\newcommand{\afchem}{Department of Chemistry, University of California, Berkeley, California 94720, USA}
\newcommand{\afbqic}{Berkeley Center for Quantum Information and Computation, Berkeley, California 94720, USA}
\newcommand{\afnottingham}{Advanced Optics and Photonics Lab, Department of Engineering, Nottingham Trent University, Nottingham NG11 8NS, UK}

\begin{document}
\preprint{APS/123-QED}

\title{Error Mitigated Metasurface-Based Randomized Measurement Schemes}

\author{Hang Ren}\thanks{Equal contributions.}
\affiliation{\afbqic}
\author{Yipei Zhang}\thanks{Equal contributions.}
\affiliation{\afbqic}
\affiliation{\afchem}
\author{Ze Zheng}\thanks{Equal contributions.}
\affiliation{\afnottingham}
\author{Cuifeng Ying}
\affiliation{\afnottingham}
\author{Lei Xu}
\affiliation{\afnottingham}
\author{Mohsen Rahmani}
\affiliation{\afnottingham}
\author{K. Birgitta Whaley}
\email{whaley@berkeley.edu}
\affiliation{\afbqic}
\affiliation{\afchem}
\date{\today}

\begin{abstract}
Estimating properties of quantum states via randomized measurements has become a significant part of quantum information science. In this paper, we design an innovative approach leveraging metasurfaces to perform randomized measurements on photonic qubits, together with error mitigation techniques that suppress realistic metasurface measurement noise. Through fidelity and purity estimation, we confirm the capability of metasurfaces to implement randomized measurements and the unbiased nature of our error-mitigated estimator. Our findings show the potential of metasurface-based randomized measurement schemes in achieving robust and resource-efficient estimation of quantum state properties.
\end{abstract}

\maketitle
\section{Introduction}

Quantum processors \cite{feyman} exploit the principles of quantum entanglement and quantum superposition to solve certain problems more efficiently than classical computers \cite{365700}. As these quantum processors scale up, characterizing the prepared quantum states becomes significantly challenging. Recently, protocols based on randomized measurements have been developed \cite{elben2023randomized},  one example of which is the method of classical shadows  \cite{huang2020predicting}. These protocols, which typically involve repeated measurements on multiple copies of the same quantum state in random bases, have demonstrated high efficacy in predicting the properties of quantum states with considerably lower resource requirements than quantum state tomography. However, the implementation of randomized measurements for photonic qubits \cite{PhysRevLett.127.200501, PRXQuantum.2.010307} requires the reconfiguration of the optical setup to modify the bases of measurements, thereby impeding the practical scalability of this approach. 

We have found that metasurfaces, two-dimensional surfaces composed of periodic sub-wavelength structured elements operating as order-selective diffractive grating \cite{wang2018quantum, zheng2023advances}, can be designed to project photons in random bases of polarization. The metasurface comprises several metagratings, each responsible for diffracting a unique pair of specific polarization states in two directions. The working principle relies on the linear dependence of the geometric phase in metagratings. This allows for manipulating wavefronts using confined electromagnetic fields, which are based on plasmonic and Mie resonances \cite{chen2016review}. When metagratings associated with multiple polarization basis pairs are concatenated, they form an equivalent polarization-dependent diffractive grating. The metasurface directs photons to spatially separated locations depending on their polarization. This process naturally enables the randomized measurements of polarization-encoded photonic qubits. Notably, a single metasurface is capable of measuring properties of quantum states comprising an arbitrary number of qubits by sequentially projecting photons, demonstrating the scalability of our approach.
In this work, we design a metasurface capable of conducting such randomized measurements and demonstrate its efficacy through numerical simulation of its optical and measurement properties.

While the theoretical analysis shows that metasurfaces can execute noiseless randomized measurements, in practice, real-world implementations are susceptible to noise originating from design and fabrication constraints that can result in biased measurement outcomes. To eliminate this noise-induced bias in estimating properties of quantum states, we analyze the physical origin of metasurface noise and construct a model that effectively explains the impact of noise on the measurement outcomes. 
The noise model is first learned through metasurface calibration. Using the calibrated noise model, we develop an error mitigation technique capable of extracting true results from noisy outcomes via subsequent post-processing. 

We have validated our protocol by performing numerical simulations of metasurfaces in the context of randomized measurements and estimation of state fidelity and state purity using classical shadows and statistical correlations of measurement outcomes \cite{huang2020predicting}. Our results indicate that the protocol effectively mitigates the impact of noise and provides accurate estimations of quantum state properties.

In contrast to existing randomized measurement approaches for photonic qubits \cite{wyderka2023complete}, our protocol avoids the need for optical setup reconfiguration. It provides a route to scalability for larger numbers of qubits. Furthermore, the proposed noise calibration and error mitigation techniques efficiently address basis-dependent and photon loss noise, requiring only moderate experimental resources.

\begin{figure*} 
\centering 
\begin{subfigure}{.35\textwidth}
  \centering
  \includegraphics[width=5cm]{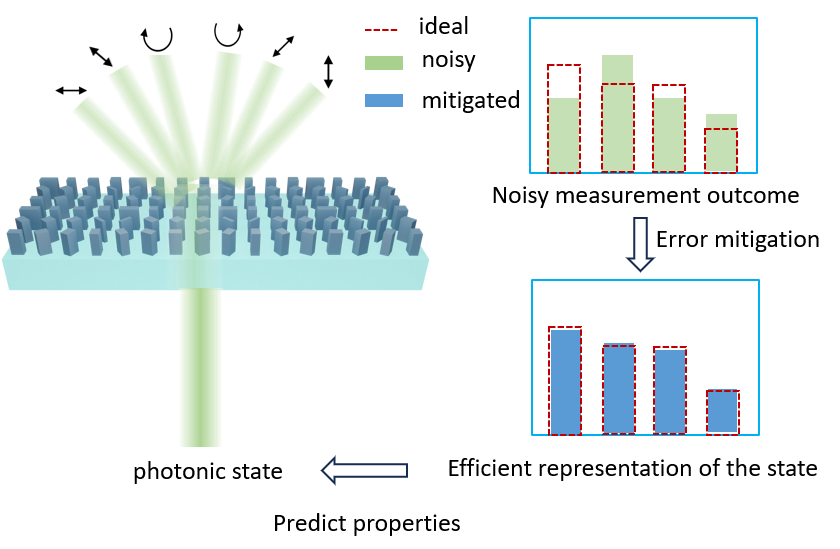}
  \caption{Error mitigated randomized measurements using a metasurface}
  \label{ms1}
\end{subfigure}%
\begin{subfigure}{.33\textwidth}
  \centering
  \includegraphics[width=4.2cm]{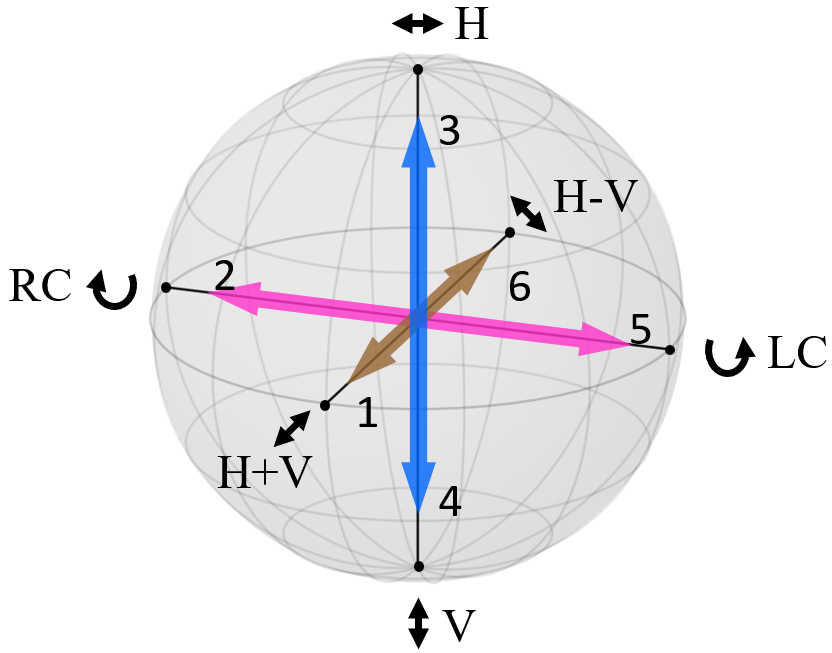}
  \caption{Measurement bases visualized in the Bloch sphere}
  \label{ms2}
\end{subfigure}
\begin{subfigure}{.3\textwidth}
  \centering
  \includegraphics[width=4.5cm]{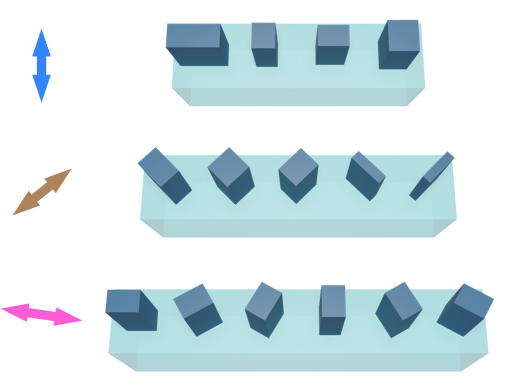}
  \caption{Periodic elements of metasurface}
  \label{ms3}
\end{subfigure}
\caption{Implementation of randomized measurements of photonic states using a metasurface. (a) The metasurface projects photons to one of six ports based on one of three randomly selected polarization basis pairs (H/V, H±V, LC/RC). H and V denote horizontal and vertical, LC and RC left circular and right circular, respectively. This process constitutes a tomographically complete set of measurements. Since the measurement of input photonic states yields noisy outcomes due to intrinsic metasurface noise, the weights of these outcomes are adjusted according to a calibrated noise model for the metasurface. This allows the error-mitigated outcomes to approximate ideal outcomes, thereby efficiently encoding state information and reducing noise-induced bias in state property predictions. (b) The Positive Operator-Valued Measure (POVM) associated with a six polarization component metasurface is represented by the six measurement projectors $ 1/3\{|x^{\pm}\rangle\langle x^{\pm}|, |y^{\pm}\rangle\langle y^{\pm} |, |z^{\pm}\rangle\langle z^{\pm} |\}$ illustrated in the Bloch sphere. (c) The measurement corresponding to each pair of POVMs is realized using a metagrating layer functioning as a binary polarization-dependent diffractive grating.\cite{zhao2022controllable}}
\label{ms_all}
\end{figure*}

The remainder of the paper is organized as follows. In Section \ref{Randomized measurements with metasurface}, we discuss the ability of metasurfaces to perform randomized measurements. Section \ref{noise modeling} details our noise model, and Section \ref{calibration} outlines the noise calibration procedures. Section \ref{error mitigation} introduces our error mitigation technique. In Section \ref{numerical results}, we numerically implement the protocol for estimating properties of quantum states,  specifically for estimating fidelity and purity of a five-qubit photonic state via the generation of classical shadows.  Finally, Section  \ref{discussion} summarizes our primary findings and suggests future research directions.

\section{Randomized measurements with metasurface}\label{Randomized measurements with metasurface}

We begin by detailing the implementation of randomized measurements using metasurfaces, employing the formalism of the Positive Operator-Valued Measure (POVM) \cite{nielsen2010quantum}. For a measurement involving $n$-qubits, the metasurface-associated POVM, denoted by the set of effects $\mathbf{E^{\mathrm{global}}}$, takes the form $\mathbf{E^{\mathrm{global}}} = \mathbf{E}^1 \times \cdots \times \mathbf{E}^q \times \cdots \times \mathbf{E}^n$. Here, each local set of effects $\mathbf{E}^q$ (where $q \in [1, n]$) represents a POVM with $K$ positive-semidefinite effects that act only on the $q^{\mathrm{th}}$ qubit. In metasurfaces, $K$ is the number of ports, which will be explained later. Furthermore, $\mathbf{E}^q \equiv \mathbf{E}$, i.e., the metasurface acts the same on all qubits. 
A detailed explanation of the working principle of the metasurface when modeled as a measurement apparatus using POVMs can be found in Appendix \ref{appendix_metasurface}.

The local set $\mathbf{E}$ comprises paired effects in the metasurface setup. This means that for any effect $E_{(i,b)}$ in the set, there is another unique effect $E_{(i,\bar{b})}$, such that $E_{(i,b)} + E_{(i,\bar{b})} = \frac{2}{K} I $. Here, $i$  denotes the measurement basis,  while $b$ and $\bar{b}$ denote the two possible measurement outcomes. One can check that $\mathbf{E}$ is indeed a POVM with $\sum_{i,b} E_{(i,b)} = I$. Each local POVM effect $E_{(i,b)}$ is proportional to a rank-one projector $E_{(i,b)} = \frac{2}{K} |\phi_{(i,b)} \rangle \langle \phi_{(i,b)} |$, and the set of these should be tomographically-complete. For the randomized measurement schemes to be practical, the pure states $|\phi_{(i,b)}\rangle$ corresponding to effects in $\mathbf{E}$ need to form a quantum 2-design \cite{gross2007evenly} (see Appendix \ref{2design} for additional details). Given these assumptions, we can write $\mathbf{E}$ as $\{E_{(i,0)}, E_{(i,1)} \}_{i = 1}^{K/2}$. In the metasurface used in this work, we have three pairs of POVM effects for each qubit, i.e., $\mathbf{E} = \{(E_{x,0}, E_{x,1}), (E_{y,0}, E_{y,1}), (E_{z,0}, E_{z,1})\}$, so that $K = 6$, where the POVM implemented by the metasurface is equivalent to the random Pauli measurement.

With the above conditions on $\mathbf{E}$, the metasurface can perform two outcome measurements in $K/2$ distinct bases. Each grating on the metasurface corresponds to a pair of POVM effects ${ E_{(i,0)}, E_{(i,1)} }$. Photons with different measurement readouts would be emitted in various directions from the grating and detected at spatially separated ports. Fig.  \ref{ms_all} illustrates the operation for $K = 6$, using the three measurement bases H/V ($z$), H $\pm$ V ($x$), LC/RC ($y$), where H, V denote horizontal and vertical polarization, and LC, RC denote left circular and right circular polarization. For example, the measurements in basis $x$ will have two possible outcomes $x^+ = $ H + V and $x^- = $H - V, labeled as $0$ and $1$, respectively.

Our task is to evaluate the functions of a quantum system characterized by an $n$-qubit density matrix $\rho$.  The first step in evaluating the functions of such a quantum system involves data collection from randomized measurements \cite{nguyen2022optimizing}. Specifically, we perform measurements defined by the POVM $\mathbf{E}^{\mathrm{global}}$ on $\rho$. The outcome $\mathbf{r} = [r^1, \dots , r^q \dots, r^n ]$ is obtained with probability $P_{\mathbf{r}}$, according to Born's rule: 

\begin{equation}\label{bornrule}
    P_{\mathbf{r}} = \mathrm{Tr}(\rho \cdot E_{r^1} \otimes \cdots \otimes E_{r^n} )
\end{equation}

The outcome for the $q^{\mathrm{th}}$ qubit, $r^q$, is a tuple $(i^q, b^q )$, where $i^q \in [1, K/2]$ represents the measurement basis, and the bit $b^q \in \{0, 1 \}$ represents the corresponding readout. This measurement process is repeated $N$ times,  yielding the data set $\{\mathbf{r}_m \}_{m=1}^N$. Following the measurements, the values of a specific function are evaluated by performing classical post-processing of the collected data, with the specific post-processing scheme dependent on the functions to be evaluated (see, e.g., \cite{huang2020predicting, elben2019statistical, brydges2019probing}). We shall discuss several specific post-processing schemes for properties of photonic qubits in detail in Section \ref{numerical results}.

\section{Noise modelling}\label{noise modeling}

The above discussion assumed that the metasurface performs ideal noise-free randomized measurements. However, accounting for the inherent sources of measurement noise in real metasurfaces is crucial. We model these noises here as general operations acting on the POVM effects. Specifically, we propose an effective noise operation $\hat{G}$, which comprises a linear transformation, denoted by a transition matrix $A$ \cite{maciejewski2020mitigation, geller2020rigorous}, followed by a nonlinear operation $\Lambda$, which depends on the POVM set $\bf{E}$ as well as on the input state $\rho$. We will see later that the noise operations $A$ and $\Lambda$ have different sources. $A$ comes from a linear transformation of the POVM set $\mathbf{E}$, while $\Lambda$ is a non-linear operator describing the result of photon loss. Thus, the effective noise operator is $\hat{G} = \Lambda \circ A$. This model accommodates a wide range of noise types, and its effectiveness will be validated in Section \ref{calibration}. For now, we shall continue with a discussion of the general structure of the proposed noise model.

Since the dominant noise in $n$-qubit measurements conducted by metasurfaces is uncorrelated, the $K^n$-dimensional global transition matrix $A$ for the linear transformation decomposes into a tensor product form \cite{maciejewski2020mitigation},
\begin{equation}
A = \bigotimes_{q = 1}^n \Gamma^q,
\end{equation}
where $\Gamma^q = \Gamma$ for all $q \in [1,n]$ represents the $K$-dimensional transition matrix for the local POVM $\mathbf{E}$ of the $q^{\mathrm{th}}$ qubit. This transforms the original POVM elements $E_{(i,b)}$ into noisy POVM effects $\tilde{E}_{(i,b)}$, according to the transition matrix $\Gamma$
\begin{equation}
\tilde{E}{(i,b)} = \sum_{(i',b')}\Gamma_{(i,b),(i',b')} E_{(i',b')} \label{linear model}.
\end{equation}
Here, the matrix element $\Gamma_{(i,b),(i',b')}$ is the probability of obtaining the measurement outcome $(i,b)$, given the idealized outcome $(i',b')$ in the absence of noise. To ensure that the generated set ${\tilde{E}_{(i,b)} }$ also constitutes a POVM, each column of $\Gamma$ must sum to one.

Metasurfaces inherently exhibit noise from factors such as the discontinuities in finite nanodisks, boundary constraints imposed by their limited size, and other inevitable design imperfections. Such noise is present even when the most precise fabrication methods are used and constitutes a significant challenge that needs to be addressed. Because the dominant intrinsic noise associated with metasurfaces does not induce transitions between different bases, i.e., no coherent noise contributes to the metasurface measurements, the transition matrix $\Gamma$ is block-diagonal, $\Gamma_{(i,b),(i',b')} = \delta_{i,i'} \Gamma_{(i,b),(i',b')}$, allowing it to be further decomposed into a direct sum form
\begin{equation}
    \Gamma=\bigoplus_{i=1}^{K / 2} \Gamma_i , \label{direct sum}
\end{equation}
where $\Gamma_i$ is the two-dimensional transition matrix that acts only on the pair of POVM effects $\{E_{(i,0)}, E_{(i,1)} \}$.

Applying Born's rule, the noisy probability distribution $\tilde{P}_{(\mathbf{i}, \mathbf{b})}$ of measurement outcomes is then related to the noise-free probability distribution $P_{(\mathbf{i},\mathbf{b})}$ via the action of transition matrix $A = \Gamma ^{\otimes n}$,
\begin{equation}
    \tilde{P}_{(\mathbf{i}, \mathbf{b})}=\sum_{\mathbf{i}^{\prime}, \mathbf{b}^{\prime}} A_{(\mathbf{i}, \mathbf{b}),\left(\mathbf{i}^{\prime}, \mathbf{b}^{\prime}\right)} P_{(\mathbf{i'}, \mathbf{b'})},
\end{equation}
where $\mathbf{i} = [i^1, \dots, i^q, \dots, i^n]$ and $\mathbf{b} = [b^1, \dots, b^q, \dots, b^n]$, with $i^q \in [1, K/2]$ and $b^q \in \{0, 1\}$ for all qubit indexes $q \in [1,n]$.

The nonlinear action $\Lambda$ depends on the noise-free POVM set $\mathbf{E}$ and the input state $\rho$. It modifies the noisy probability distribution $\tilde{P}_{(\mathbf{i}, \mathbf{b})}$, in a way that depends nonlinearly on all the noise-free probabilities $\{ P_{(\mathbf{i}, \mathbf{b})}\}$ that derive from the input state. The mathematical representation of the action of $\Lambda$ will be discussed in Section \ref{subsec:photon}.

We now move to a detailed examination of the specific intrinsic noise present in metasurfaces and its physical origin. It is important to note that the noise models we consider here are not due to defects or imperfect experimental settings but are inherited from the design of the metasurfaces. We first address linear noise sources, i.e., the effect of $A$, which derives here from finite size and imperfect material design of the metagratings (Sections \ref{subsec:bitflip} and \ref{subsec:amplitude}). 
We then address the nonlinear noise $\Lambda$ due to photon loss in Section \ref{subsec:photon}. In Section \ref{subsec:composite}, we employ an effective noise model that describes the combined influence of the aforementioned noise sources.

\subsection{Basis-flip and depolarizing noise} \label{subsec:bitflip}

A metasurface consists of a finite number of discrete units. Due to the phase change discontinuity and boundary conditions arising from its finite size, readout errors can occur. The construction of traditional optical gratings relies on continuous gradual phase changes accumulated along the optical path. On the contrary, metagratings introduce a distribution of the abrupt phase shift caused by each subwavelength unit. The limited unit numbers provide discrete phase distribution, affecting the efficiency of the polarisation-dependent diffraction. \cite{ma14092212}

The most common type of measurement noise resulting from the phase change discontinuity is due to the finite size of the metasurface. The resulting finite size effect on photonic qubit readout causes a flip in the measured basis states, e.g., when a horizontally polarized photon is erroneously diffracted in a direction associated with vertical polarization. Throughout this work, we use basis-flip noise to refer to this basis-dependent noise causing basis-state flips.
 
We take the basis-flip noise to occur with a probability $p_{bf}(i)$ for the $i$th pair of POVM effects.  The outcome $b$ associated with the POVM effect $E_{(i,b)}$ is correspondingly flipped to its paired outcome $\bar{b}$ associated with the POVM effect $E_{(i,\bar{b})}$ \cite{smith2021qubit}. 
In Fig. \ref{ms_all}, we show the effects of random Pauli measurements when K equals 6. Here, the specific basis-flip noise becomes a more general form of Pauli noise. For instance, consider the basis-flip noise during a z-basis measurement, where there is a possibility of misreading H (horizontal) polarization as V (vertical) polarization. This type of error results from the conventional bit-flip noise, also known as X errors, and is compounded by Y errors. Consequently, the total basis-flip rate for z-basis measurements is determined by combining the rates of both X and Y errors. Similarly, errors in both X and Z measurements contribute to basis state flips during y-basis measurements. We also consider depolarizing errors, where 
with a certain probability $p_d$, the photon is output with a uniformly random probability of $1/K $ from all detection ports.\cite{nguyen2022optimizing}.

These types of measurement noise derive from diverse physical sources. For instance, depolarizing noise can arise from a finite coupling strength between the photon and the measurement device, resulting in weak (non-projective) measurement  \cite{oreshkov2005weak, tamir2013introduction}. In contrast, basis-flip noise may be attributed to the open boundary and discrete nature of the metasurfaces \cite{wu2018modelling}.

The action of each of these types of measurement noise for the pairs of local POVM effects $\{E_{(i,b)}, E_{(i,\bar{b})} \}$ is described by a channel. For example, we have $K/2$ basis-flip channels acting on local POVM effects where each channel contains two complementary effects of the local POVM. These $K/2$ channels have the same form but different parameters. More precisely, the impact of the basis-flip noise on basis $i$ with the rate $p_{bf}(i)$ is given by $\tilde{E}_{(i,b)}= \mathcal{E}_{bf}\left(E_{(i,b)}\right)=p_{bf}(i) \cdot E_{(i,\bar{b})}+\left(1- p_{bf}(i)\right) \cdot E_{(i,b)} = 2 p_{bf}(i) \cdot \frac{I}{K}+\left(1-2 p_{bf}(i)\right) \cdot E_{(i,b)}$, where we have used the pairing condition $E_{(i,b)} + E_{(i,\bar{b})} = \frac{2I}{K}$ for basis $i$. Similarly, the impact of depolarizing noise on basis $i$ with rate $p_d$ is given by the channel $\tilde{E}_{(i,b)} = \mathcal{E}_{d}\left(E_{(i,b)}\right)=p_{d} \cdot \frac{I}{K}+\left(1- p_{d}\right) \cdot E_{(i,b)}$ \cite{nguyen2022optimizing}. Due to the same form of $\mathcal{E}_{bf}$ and $\mathcal{E}_{d}$, we can merge these two types of channels and use a single noise rate $p_{bf}(i)$ for each basis $i$ to describe their impacts on the probability distribution associated with the POVM effects for mathematical convenience. In this case, the transition matrix   $\Gamma^{bf}_i$ that corresponds to the $i^{\mathrm{th}}$ basis is then
\begin{equation}\label{eq:bf_transition_matrix}
\Gamma^{bf}_i=\left[
\begin{array}{cc}
1-p_{bf}(i) & p_{bf}(i) \\
p_{bf}(i) & 1-p_{bf}(i)
\end{array}
\right],
\end{equation}
where $p_{bf}(i)$ is the bit-flip error rate for the $i^{\mathrm{th}}$ basis, and for example, $i=1,2,3$ correspond to H/V, H $\pm$ V, and LC/RC in Fig. \ref{ms_all}. Eq. \eqref{eq:bf_transition_matrix} describes the combined and indistinguishable impacts of the basis-flip and depolarizing noises. From now on, we call this noise effective basis-flip noise.

\subsection{Amplitude damping noise} \label{subsec:amplitude}

Suppose a metasurface is positioned in the horizontal plane and photons are projected vertically towards the metasurface (see Fig. \ref{ms1}); the photon wavefront spreads across the metasurface. However, since the metasurface exhibits asymmetry in the horizontal plane, the resulting diffraction pattern depends on the pattern of the individual metagratings composing the metasurface. This can lead to unevenly distributed measurement noise for a particular pair of polarizations. For instance, in Fig. \ref{ms_all}, while the H polarization might be measured with high fidelity, the V  polarization could be more susceptible to being misread as H. Similar situations arise for measurements in other bases as well. Such noise behavior can be represented by photons passing through an amplitude-damping quantum channel in the $i^{\mathrm{th}}$ basis.
Thus, in the context of photon measurement by the metasurface, we have amplitude damping noise when a photon that should ideally be projected to $E_{(i,1)}$ is instead projected to $E_{(i,0)}$ with probability $p_{ad}(i)$. This error can be seen as a result of an unwanted asymmetry in the design of the $K/2$ pairs of polarization ports.

The amplitude damping noise on the POVM effects pair $\{ E_{(i,1)}, E_{(i,0)} \}$ can be described by a
channel \cite{nielsen2010quantum} such that $\tilde{E}_{(i,0)}=\mathcal{E}_{ad}(E_{(i,0)})=E_{(i,0)}+p_{a d}(i) E_{(i,1)}$
and $\tilde{E}_{(i,1)}=\mathcal{E}_{ad}(E_{(i,1)})=\left(1-p_{a d}(i)\right) E_{(i,1)}$, and the associated transition matrix $\Gamma^{ad}_i$ can be derived as 
\begin{equation}
    \Gamma^{ad}_i=\left[\begin{array}{cc}
1 & p_{a d}(i) \\
0 & 1-p_{a d}(i)
\end{array}\right],
\end{equation}
where $p_{ad}(i)$ is the amplitude damping error value for the $i^{\mathrm{th}}$ basis.

When combined with the aforementioned basis-flip noise, the overall effect of 
these two types of stochastic noise lead to asymmetric basis-flip noise between a pair $\{E_{(i,0)}, E_{(i,1)} \}$ with unequal basis-flip rates, a situation that is easily described by a 2D transition matrix $\Gamma_i$, which is the product of basis-flip and amplitude damping transition matrices $\Gamma_i = \Gamma^{ad}_i \, \Gamma^{bf}_i$.

\subsection{Photon loss noise} \label{subsec:photon}

Photon loss noise is a form of nonlinear noise, distinct from the linear transformations on POVM effects discussed so far. This type of noise can arise in several circumstances, such as when a photon is reflected due to a limited transmission rate, a photon detection port fails to detect the existence of a photon, or a photon emerges from a higher-order diffraction direction after coupling with the metasurfaces \cite{chen2016review}. 

The photon loss rates $p_{pl}(i, b)$ that are associated with each detecting port $(i, b)$ may vary, both with measurement basis $i$ and with outcome $b$. Unequal photon loss rates for a pair of effects ${E_{(i,0)} , E_{(i,1)} }$ in the POVM can then result in the output string probabilities deviating from their true values.  

To see this explicitly, consider the noise-free probabilities $P_{\mathbf{i}}=\{P_{(\mathbf{i}, \mathbf{b})}\}_{\mathbf{b}\in \{ 0,1\}^n}$ and the corresponding noisy probabilities  $\tilde{P}_{\mathbf{i}} = \{\tilde{P}_{(\mathbf{i}, \mathbf{b})}\}_{\mathbf{b}\in \{ 0,1\}^n}$ for basis $\mathbf{i} = [i^1\dots i^n]$. These two probability sets  are related by a nonlinear transformation $\Lambda$ such that \begin{equation}\label{eq: nonlinear map}
    \tilde{P}_{\mathbf{i}} = \Lambda[P_{\mathbf{i}}].
\end{equation} In particular,  for any $ \tilde{P}_{(\mathbf{i},\mathbf{b})} \in \tilde{P}_{\mathbf{i}} $, we have
\begin{equation}\label{eq:nonlinear}
    \tilde{P}_{(\mathbf{i}, \mathbf{b})} =  \frac{P_{(\mathbf{i}, \mathbf{b})}  \left[ 1-p_{pl}(\mathbf{i},\mathbf{b})\right]}{\sum_{\mathbf{b} \in \{0,1\}^n}P_{(\mathbf{i}, \mathbf{b})}  \left[ 1-p_{pl}(\mathbf{i},\mathbf{b})\right]},
\end{equation}
where $1-p_{pl}(\mathbf{i},\mathbf{b})= \prod_{q=1}^n \left[ 1-p_{pl}(i_q,b_q)\right]$, with $\mathbf{b} = [b^1\dots b^n]$ indicating the measurement outcome.

Clearly, photon loss has the greatest impact on the output statistics when the loss rates are unequal. If the loss rates across different measurement ports are the same, it implies an equal reduction in measurement efficiency for all outcomes. This scenario can be interpreted as a uniform loss of data. While this uniformity does not introduce bias in the estimation process, which means $\tilde{P}_{(\mathbf{i},\mathbf{b})} = P_{(\mathbf{i},\mathbf{b})}$ as can be seen from Eq. \eqref{eq:nonlinear}, it does lead to an increase in the variance of the estimation, thereby impacting our statistical inference.

\subsection{Composite noise model} \label{subsec:composite}

We can now summarize an effective noise model $\hat{G}$ acting on the noise-free probabilities $\{P_{(\mathbf{i}, \mathbf{b})}\}$ that combines the linear and nonlinear measurement noise processes described above. This model describes a sequence of possible transformations: first, an effective basis-flip operation, followed by an amplitude damping operation, 
and finally, a nonlinear operation due to photon loss. It is important to emphasize that the basis-flip and amplitude-damping operations are commutative.  Yet, as we shall explain later, placing the nonlinear operation $\Lambda$ after $A$ is crucial to ensure consistency between the noise model and error mitigation development. Note that our model does not consider the possibility of any class of error occurring two or more times in a given error cycle since such events have exponentially smaller probabilities. Hence, we maintain the sequence of noise operations as presented above, i.e., 
\begin{equation}\label{eq: overall_operation}
    \hat{G} = \Lambda \circ A,
\end{equation} 
where the nonlinear map $A = \left[ \bigoplus_{i=1}^{K / 2} \Gamma^{ad}_i \, \Gamma^{bf}_i \right]^{\otimes n}$, and $\Lambda$ is described in Eq. \eqref{eq: nonlinear map} and Eq. \eqref{eq:nonlinear}.
The vector of noisy probability distributions of measurement outcomes $\tilde{P} = \{\tilde{P}_{(\textbf{i},\textbf{b})}\}$ is then related to the vector of ideal noise-free distributions $P = \{ P_{(\textbf{i},\textbf{b})}\}$ by
\begin{equation} \label{eq:noisyProbs}
    \tilde{P} = \hat{G}_{\boldsymbol{\lambda}} \left[ P \right],
\end{equation}
where $\boldsymbol{\lambda} = [p_{bf}(i), \, p_{ad}(i), \, p_{pl}(i,0), \,p_{pl}(i,1)]_{i = 0}^{K/2}$ is a vector representing the noise values for multiple noise sources, and $\hat{G}_{\boldsymbol{\lambda}}$ can be regarded as a $\boldsymbol{\lambda}$-dependent noise map acting on the ideal probabilities which combines the impact of linear and nonlinear noises as per Eq. \eqref{eq: overall_operation}. $\hat{G}$ satisfies $\hat{G}_{\boldsymbol{\lambda} = \vec{\boldsymbol{0}}} = \hat{I}$, i.e. if the noise value are all $0$s, $\hat{G}$ acts as an identity map.

The impact of these noise sources is modeled by applying the relevant transformations in Eq. \eqref{eq:noisyProbs} to the noise-free probability distributions $P_{(\textbf{i},\textbf{b})}$, resulting in the set of noisy distributions $\tilde{P}_{(\textbf{i},\textbf{b})}$. The error mitigation discussed in section \ref{error mitigation} can be viewed as applying an inverse transformation to the noisy data that allows for extracting true probability distributions for measurement outcomes from the noisy photon count distributions. 

It's important to note that the error mitigation strategy is not designed to correct individual data points, as with error correction. Instead, it aims to adjust the overall statistical expectations derived from the data set.

\section{CALIBRATION OF INTRINSIC NOISE CHANNELS }\label{calibration}
\begin{figure}
    \centering
    \includegraphics[width=8.4 cm]{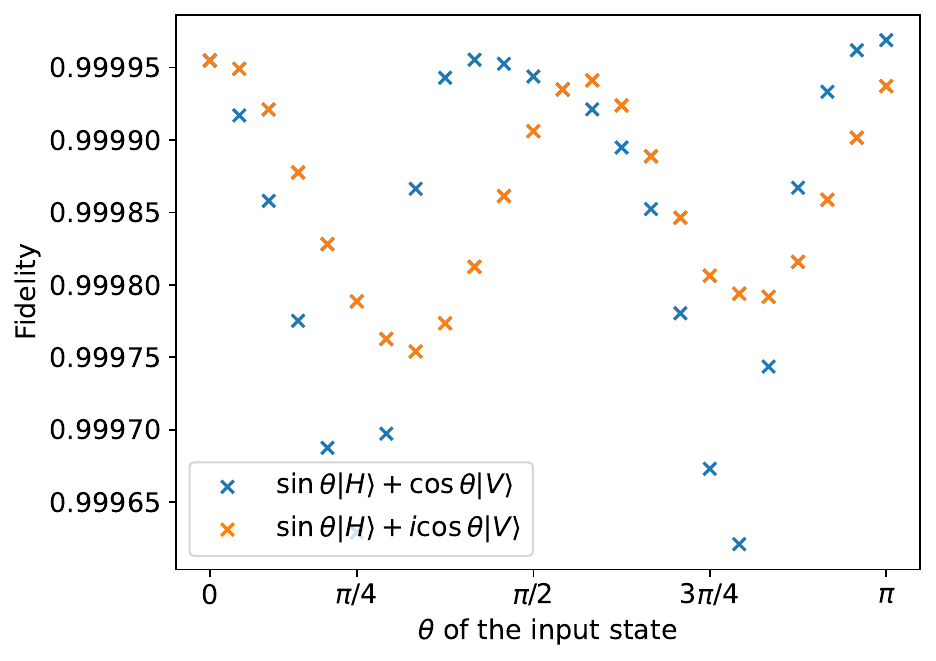}
    \caption{Evaluation of noise modeling and calibration. The calibrated noise model is utilized for prediction, with fidelity calculated against metasurfaces measurement outcomes. The evaluations are conducted for different states, each characterized by different relative amplitudes and phases. The resulting fidelity values consistently exceed 0.99, suggesting that the calibrated noise effectively models the intrinsic noise of the metasurface. This provides an empirical validation of the accuracy of the noise model constructed here.}
    \label{fig:enter-label}
\end{figure}

This section outlines the method used to design the noise calibration experiment and the protocol to determine the parameters $\boldsymbol{\lambda}$ for the noise map $\hat{G}_{\boldsymbol{\lambda}}$. In principle, one can perform a tomography-like experiment, such as quantum detector tomography, to learn all matrix elements of every noisy POVM effect $\tilde{E}_{(i,b)}$ \cite{lundeen2009tomography}. However, this task becomes highly resource-intensive as the number of qubits increases. \cite{PhysRevLett.123.140405, Shaffer2023sampleefficient}

Informed by our prior analysis of the noise origins in metasurfaces, we can adopt the noise model defined by a limited set of parameters as discussed in section \ref{noise modeling}. We can design a calibration protocol to learn this parameterized noise model with fewer experimental resources.

Specifically, for the linear noise operations, we consider uncorrelated noise models without contributions from coherent noise, i.e., the linear part of the error model takes the form of Eq. (\ref{direct sum}) so that transitions are allowed only between paired POVM effects $\{E_{(i,0)}, E_{(i,1)} \}$. The corresponding noise-induced transitions are then described by the tensor product of the Pauli stochastic noise channels of single qubit basis-flip and amplitude damping contributions detailed in Sections \ref{subsec:bitflip} - \ref{subsec:composite}.

We take the case with six ports as an example. The calibration protocol proceeds as follows. First, we use COMSOL\cite{multiphysics1998introduction} to emulate the interaction of photons with metagratings, i.e., noisy measurements using the designed metasurface composed of three different types of metagratings that are described in Appendix \ref{appendix_metasurface}. The parameters for these metagratings are summarized in Tables \ref{characterization} and \ref{calibration data}.  

This step emulates how photons, when incident on the metagratings, are diffracted to different ports, resulting in the collected photon counts. The emulation aims to replicate the physical noise in actual experiments, thus providing a realistic basis for calibration.

In this emulation, six input states ($C=6$), denoted as $\{|k \rangle \}_{k =1}^{C}$, are selected. For each state, the associated outcome probability distributions, which are influenced by noise, are emulated by COMSOL and denoted as $\{ \bar{P}(k) \}_{k = 1}^C$. These distributions reflect how the physical noise affects the photon measurement outcomes.

The central aim of the calibration is to model the impact of physical noise through a parameterized noise channel and characterize these parameters. This involves adjusting noise parameters, collectively represented as $\boldsymbol{\lambda}$, and calculating the randomized measurement outcomes accordingly, such that these outcomes align with those obtained from the COMSOL emulations. By doing so, we can determine twelve noise parameters, including three effective basis-flip, three amplitude damping, and six photon loss values. The noise model is detailed in section \ref{noise modeling}.

Concretely, given a set of noise parameters, we can calculate the six sets of outcome probability distributions from the simulation of the parameterized noise model as $\{ \tilde{P}(k, \boldsymbol{\lambda}) \}_{k = 1}^C$. This is achieved by first employing Born's rule to calculate the noise-free probabilities for each input state. Subsequently, these probabilities are subjected to a total noise map, denoted as $\hat{G}_{\boldsymbol{\lambda}}$, to obtain the noisy probabilities $\{ \tilde{P}(k, \boldsymbol{\lambda}) \}_{k = 1}^C$.

The noise calibration is identifying the noise parameters that yield simulation outcomes $\{ \bar{P}(k) \}_{k = 1}^C$, which closely resemble the emulation outcomes $\{ \tilde{P}(k, \boldsymbol{\lambda}) \}_{k = 1}^C$. The measure used to quantify the closeness of these two sets of outcomes is the Bhattacharya distance ~\cite{bhatt1946measure}, a metric for evaluating the fidelity $F_{pr}$ between two classical probability distributions $p$ and $q$ in Eq. \eqref{eq: fidelity_pq}.

\begin{equation} \label{eq: fidelity_pq}
    F_{pr}(p,q) = \left( \sum_i \sqrt{p_i q_i} \right)^2.
\end{equation}
The parameters $\boldsymbol{\lambda}$ of the noise models are obtained via the solution of the following optimization problem:
\begin{equation} \label{eq:optimize}
    \boldsymbol{\lambda}_{opt} = \mathrm{arg} \max_{0\leq\lambda_i \leq 1} \left[\sum_{k} F_{pr}\left(\bar{P}(k), \tilde{P}(k, \boldsymbol{\lambda})\right) \right].
\end{equation}

For the six-port case with POVMs visualized in Fig. \ref{ms_all}, the POVM $\mathbf{E} = 1/3\{| x^{\pm} \rangle \langle x^{\pm}|, |y^{\pm}\rangle \langle y^{\pm}|, |z^{\pm} \rangle \langle z^{\pm}|\}$, where the POVM elements are eigenstates of Pauli $X$, $Y$, and $Z$ operators, i.e., $\mathbf{E}$ is the set of effects deriving from random Pauli measurements. 

The $C = 6$ input states are $| x^{\pm} \rangle$, $|y^{\pm}\rangle$, and $|z^{\pm} \rangle$, and they produce corresponding $6$ sets of outcome probability distributions 
$\{ \tilde{P}(k, \boldsymbol{\lambda}) \}_{k = 1}^C$. The six input states and three pairs of measurement outcomes yield eighteen independent equations, which are sufficient for determining the twelve noise parameters. By solving the optimization problem in Eq. \eqref{eq:optimize}, we can obtain the optimal vector of error parameters $\boldsymbol{\lambda}_{opt}$ which are listed in Table \ref{noise rate}. 

In this calibration protocol, we model the metasurface noise with a parameterized noise model and find the noise parameters through optimization. This calibration is critical for ensuring that the model accurately reflects the real-world noise effects on photon measurements in quantum experiments.

To test the validity of the fitted error model $\hat{G}_{\boldsymbol{\lambda}_{opt}}$, we then take input states that are sampled from single-qubit superposition states in the $xz$ and $yz$ planes, i.e., $\psi_{xz}(\theta) = \mathrm{sin}(\theta)|H\rangle + \mathrm{cos}(\theta)|V\rangle $ and $\psi_{yz}(\theta) = \mathrm{sin}(\theta)|H\rangle + i\mathrm{cos}(\theta)|V\rangle $. We calculate the fidelity of the outcome probabilities $F(\bar{P}(\theta), \tilde{P}(\theta, \boldsymbol{\lambda}_{opt}))$ as a function of $\theta$, where $\tilde{P}(\theta, \boldsymbol{\lambda}_{opt})$ is obtained from our fitted model $\hat{G}_{\boldsymbol{\lambda}_{opt}}$, and $\bar{P}(\theta)$ is from simulation of the metasurface using COMSOL. The results are shown in Fig. \ref{fig:enter-label}, from which we can see the fidelity is above 0.99 for all sampled values of $\theta$. This provides empirical validation of our generic measurement noise model and our designed calibration protocol that characterizes these noises.

\section{Error mitigation}\label{error mitigation}

This section presents an error mitigation scheme to remove bias induced by the previously discussed error models. The method we present applies to any randomized measurement protocol since it operates on the level of measurement outcome probabilities. This is important because we aim to address basis-dependent measurement noise, which has previously proven challenging to handle \cite{chen2021robust, koh2022classical}. 

Regarding the linear components of the noise, once we have constructed the transition matrix $A$ via calibration, we can then apply the inverse of $A$ to infer noise-free probabilities $P_{(\mathbf{i}, \mathbf{b})}$ from the empirical estimations of the noisy probabilities $\tilde{P}_{(\mathbf{i}, \mathbf{b})}$ (denoted as $\bar{P}_{(\mathbf{i}, \mathbf{b})}$)
\begin{equation}
    P_{(\mathbf{i}, \mathbf{b})}=\sum_{\mathbf{i}^{\prime}, \mathbf{b}^{\prime}} [A]^{-1}_{(\mathbf{i}, \mathbf{b}),\left(\mathbf{i}^{\prime}, \mathbf{b}^{\prime}\right)} \bar{P}_{(\mathbf{i}^{\prime},\mathbf{b}^{\prime})}.
\end{equation}

In the above construction, the estimated noisy probabilities on the right-hand side are influenced by noise originating from the metasurface and statistical fluctuation due to the limited number of measurement samples.

In the case of photon loss noise, we can employ an analogous error mitigation scheme to infer the noise-free probabilities $P_{(\mathbf{i},\mathbf{b})}$ from the noisy empirical estimation $\bar{P}_{(\mathbf{i},\mathbf{b})}$ of the noisy probabilities $\tilde{P}_{(\mathbf{i},\mathbf{b})}$. If the photon loss rates associated with the 6 POVM effects $\{ E_{(i,b)} \}$ are $p_{pl}(i, b)$, then we can construct $P_{(\mathbf{i},\mathbf{b})}$ via the nonlinear renormalization procedure
\begin{equation}
    P_{(\mathbf{i},\mathbf{b})} = \frac{k_{(\mathbf{i},\mathbf{b})}}{\sum_{\mathbf{b}}k_{(\mathbf{i},\mathbf{b})}},
\end{equation}
where $k_{(\mathbf{i},\mathbf{b})}$ is obtained from $\bar{P}_{(\mathbf{i},\mathbf{b})}$ as
\begin{equation}
    k_{(\mathbf{i},\mathbf{b})} = \frac{\bar{P}_{(\mathbf{i},\mathbf{b})}}{\prod_{q = 1}^{n} [1 - p_{pl}(i_q, b_q)]}.
\end{equation}

Here $k_{(\mathbf{i},\mathbf{b})}$ gives the renormalization of the empirical measurement outcomes $\bar{P}_{(\mathbf{i},\mathbf{b})}$ by the photon loss probabilities for each basis of the metasurface measurement. One can immediately notice from the above equation that if $p_{pl}(i, 0) = p_{pl}(i, 1)$ for $\forall i$,
 then photon loss error would not affect the output string probability, i.e., $\tilde{P}_{(\mathbf{i},\mathbf{b})} = P_{(\mathbf{i},\mathbf{b})}$. However, the material design is generally more likely to give unequal probabilities.

\begin{figure}
    \centering
    \includegraphics[width=8.6 cm]{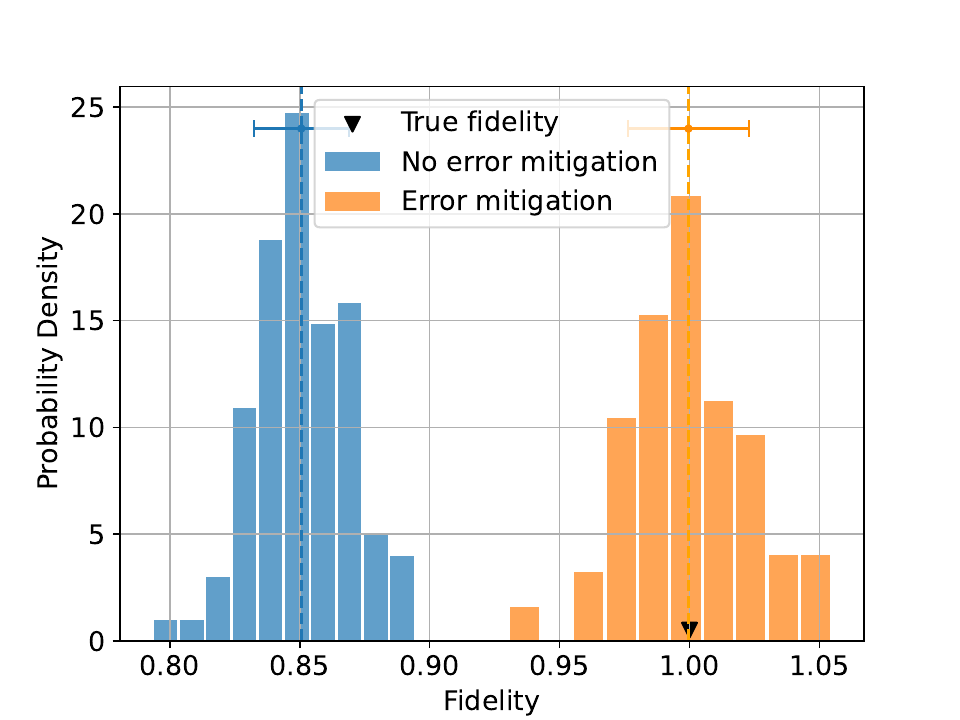}
    \caption{Estimated fidelity distribution of a five-qubit W state using classical shadows. The fidelity is estimated from $10^4$ metasurface measurements of $\rho$, i.e., $10^4$ values of $m$ in Eq. \ref{eq:shadow_k}. Error-mitigated randomized measurements lead to unbiased estimation of fidelity (true fidelity = 1).}
    \label{distribution}
\end{figure}

\section{Evaluation of Protocol Performance: Estimating Measurement Fidelity and Purity}\label{numerical results}

\begin{figure*}
\centering
\begin{subfigure}{.5\textwidth}
  \centering
  \includegraphics[width=8.6cm]{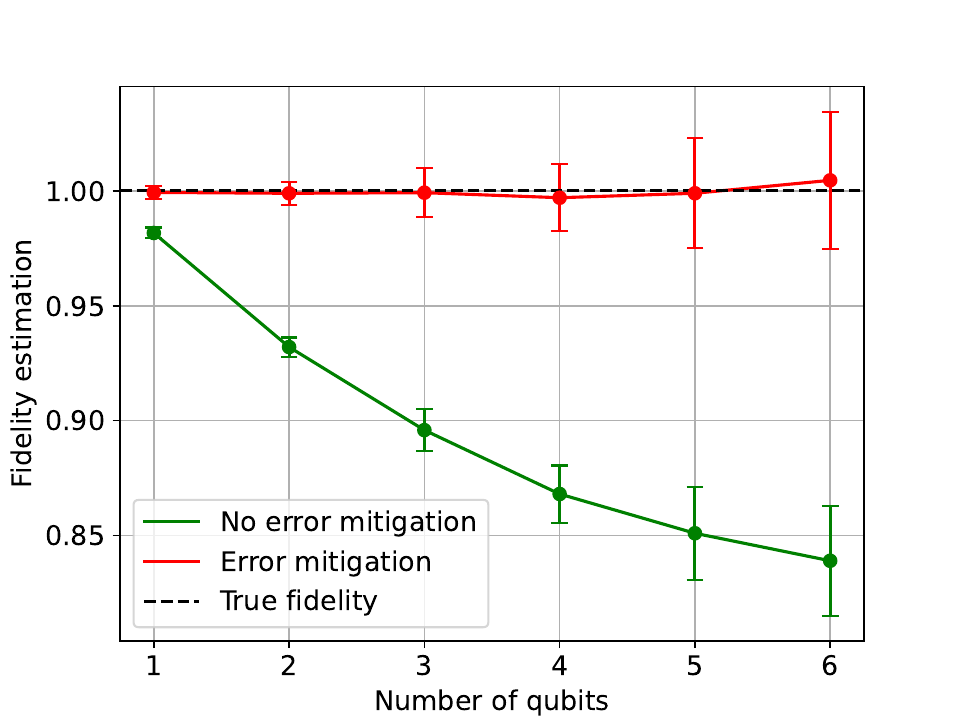}
  \caption{Fidelity of W states with different numbers of qubits.}
  \label{}
\end{subfigure}%
\begin{subfigure}{.5\textwidth}
  \centering
  \includegraphics[width=8.6cm]{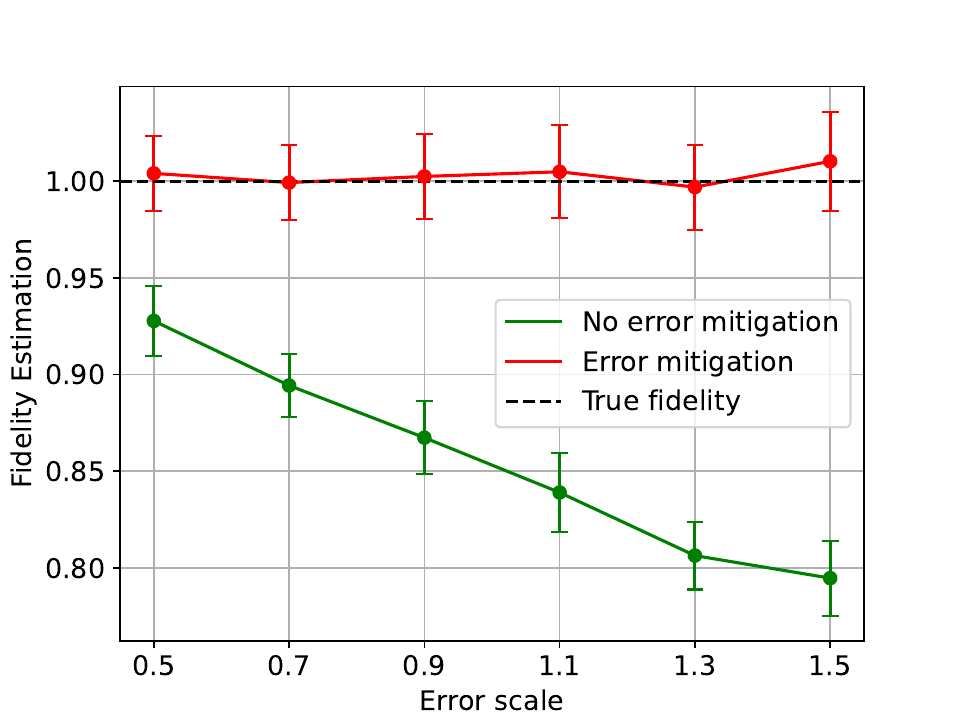}
  \caption{Fidelity of five qubit W state with adjusted noise levels.}
  \label{}
\end{subfigure}
\caption{Measurement fidelity estimation of W states using classical shadows. The fidelity is evaluated by taking the average over $10^4$ classical shadows according to Eq. \ref{eq:measureF}. 
Increased system size or higher error rates increase the noise effect on the final state, decreasing fidelity. However, the noise-induced bias in fidelity estimation can be effectively removed with the error mitigation protocol. Overall, the protocol shows scalability relative to (a) the number of qubits and (b) exhibits stability within a broad range of error scale $h = |\boldsymbol{\lambda^\prime}|/|\boldsymbol{\lambda}|$.}
\label{fidelity}
\end{figure*}

\begin{figure*}
\centering
\begin{subfigure}{.5\textwidth}
  \centering
  \includegraphics[width=8.6cm]{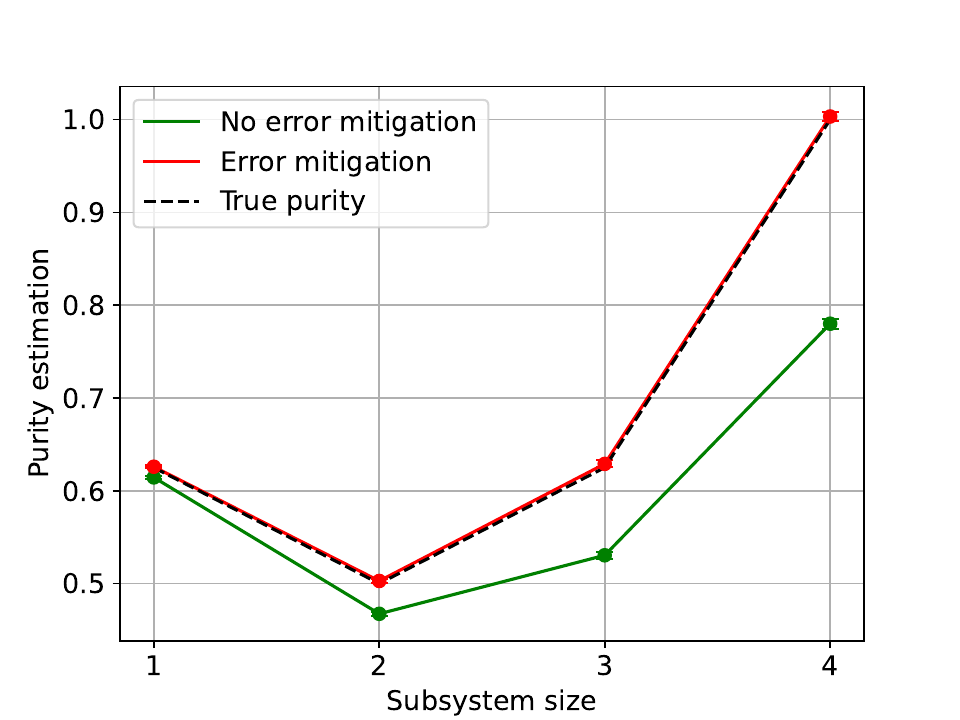}
  \caption{Purity of W states for four qubits measured with different subsystem sizes.}
  \label{}
\end{subfigure}%
\begin{subfigure}{.5\textwidth}
  \centering
  \includegraphics[width=8.6cm]{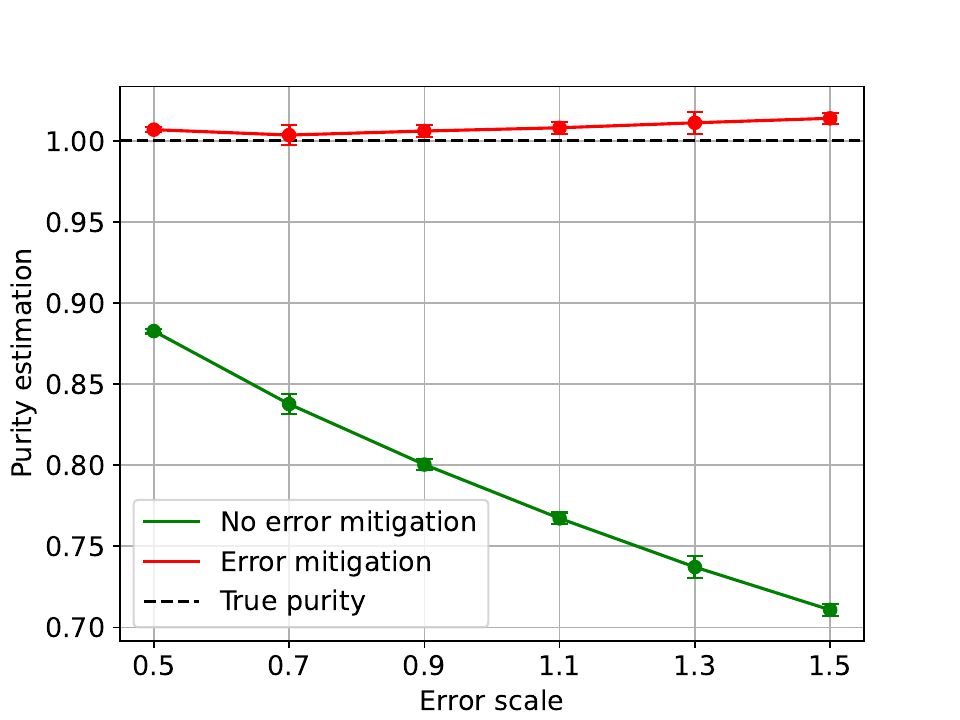}
  \caption{Purity of W states for four qubits measured under adjusted noise levels.}
  \label{}
\end{subfigure}
\caption{Purity estimation of a four-qubit W state using metasurface-based randomized measurements. The purity is calculated as the weighted hamming distance between the measured outcomes. The purity of a subsystem (reduced density matrix) can quantify the entanglement of the entire system. Panel (a) shows that the use of error-mitigated measurement outcomes results in an unbiased estimation of subsystem purity. Panel (b) demonstrates the efficacy of the protocol across a broad noise regime.
}
\label{purity}
\end{figure*}
We now evaluate the capability of metasurfaces to perform randomized measurements and the effectiveness of the error mitigation protocol described above. We shall focus on the properties of the n-qubit  W states for the photonic system:
\begin{equation}
|w\rangle = (|10\ldots 0\rangle + |01\ldots0\rangle + |00\ldots1\rangle)/\sqrt{n}.
\end{equation}

These states are commonly used in quantum teleportation \cite{dur2000three, pradhan2008generalized}, which typically employs photons as qubits. W states are thus applicable target states for analysis of the performance of our protocol.

Randomized measurements utilizing the metasurfaces are performed on the photonic W states, yielding noisy outcomes that are collected for error mitigation. The error-mitigated outcomes undergo post-processing to extract the desired quantum state property. 

We first consider the estimation of the density matrix by the method of classical shadows~\cite{huang2020predicting}. In this approach, the estimator $\hat{\rho}$ of the original density matrix $\rho$ is given by 
\begin{equation} \label{eq:shadow_k}
    \hat{\rho}_{m} = \hat{M}^{-1}(E_{\mathbf{r}_m}),
\end{equation}
where the reconstruction map $\hat{M}^{-1} = \hat{M}^{-1}_1 \otimes \hat{M}^{-1}_2 \otimes \cdots \otimes \hat{M}^{-1}_n $ with $\hat{M}^{-1}_i(X) = 3 X - I$. 

$\hat{\rho}_m$ is referred as a classical shadow from the $m^{\mathrm{th}}$ measurement outcome $\mathbf{r}_m$, as in Eq. \ref{bornrule}. The classical shadows efficiently store information about the state so that it can be easily stored on a classical computer. One can verify that the classical shadow is an unbiased estimator of the original density matrix, in the sense that classical shadows averaged according to the probability distribution $P_{\mathbf{r}} = P_{(\mathbf{i},\mathbf{b})}$ in Eq. \eqref{bornrule} 
 converge to the original density matrix, i.e., $\mathbb{E}[\hat{\rho}] = \lim_{N \rightarrow \infty } 1/N \sum_m \hat{\rho}_m = \rho$. For an observable $\hat{O}$, the estimation of its expectation value given the density matrix $\rho$ is then calculated from the explicit expression $\mathrm{Tr}(\hat{O} \rho) = 1/N \sum_m \mathrm{Tr}( \hat{O} \hat{\rho}_m)$. For the estimation of fidelity with the W state, we use
\begin{equation} \label{eq:measureF}
    \mathbf{F}(\rho, |W\rangle\langle W|) =\mathbb{E}_m[\mathrm{Tr}(|W\rangle\langle W|\hat{\rho}_m)]. 
\end{equation}

We first employ this protocol using the error calibration, measurements of photonic states, and error mitigation described in Sections \ref{calibration} and \ref{error mitigation} to numerically evaluate the measurement fidelity of a five-qubit W state, utilizing ten thousand classical shadows collected through randomized measurements via a metasurface. This entire process is repeated one hundred times. The results for both noisy and error-mitigated fidelity distributions are graphically represented in Fig. \ref{distribution}. As might be expected, the presence of measurement errors deriving from the intrinsic noise sources of the metasurface leads to significant deviations in fidelity from the true value of one, indicating that the noisy results do not accurately represent the state information. In contrast, the error-mitigated fidelity distribution is now centered around the true value, demonstrating the effectiveness of our error mitigation protocol in rectifying measurement outcomes and effectively removing measurement bias stemming from noise.

We observe that the fidelity can occasionally exceed one because we rely on finite samples, which might result in an unphysical density matrix. However, the distance metric that assesses how close the value is to one continues to be a valid measure of fidelity. This principle also extends to the purity estimation described below. To guarantee that both fidelity and purity remain below one, physical constraints can be incorporated into the data processing using the maximum likelihood estimation method \cite{MENG2023106661}, which is beyond the scope of this paper. In this study, we employ the standard data-processing method of randomized measurements to avoid inconsistencies and bias in the results.

We are also interested in evaluating the performance of the protocol across different system sizes and under various error regimes\cite{cheng2023fidelity}. In Fig. \ref{fidelity}, we utilize $10^4$ classical shadows, each generated from a single photon count, to estimate the fidelity of the noisy and error-mitigated states concerning the true W state, with panel (a) showing the dependence on the number of qubits in the W states, and panel (b) showing the dependence on the metasurface error scale $h = |\boldsymbol{\lambda^\prime}|/|\boldsymbol{\lambda}_{opt}|$, where $\boldsymbol{\lambda}_{opt}$ corresponds to the noise parameter characterized from the designed metasurface, and $ \boldsymbol{\lambda^\prime}$ denotes the re-scaled noise parameters we obtained from $\boldsymbol{\lambda}_{opt}$ by multiplying a factor $h$, i.e., $\boldsymbol{\lambda^\prime} = h \boldsymbol{\lambda}_{opt}$. Here $|\cdot|$ represents the magnitude of the vector. Remarkably, we find that across a wide range of system sizes and errors, the error-mitigated method of classical shadows predicts a fidelity close to one, indicating that our error mitigation protocol offers an unbiased estimator of state properties in the presence of noise.

In addition to utilizing classical shadows, properties of $\rho$ can be directly estimated from the collection of output strings $\{\mathbf{r}_m \}_{m = 1}^{N}$, where $\mathbf{r}_m$ is the outcome of the $m^{\mathrm{th}}$ measurement defined in Eq. \eqref{bornrule}. One such property of significance for understanding the complexity of quantum states and how they can be used to process quantum information is the degree of entanglement \cite{PhysRevLett.87.050401, hu2024demonstration}.

This enables easy access to another quantitative estimate of the measurement error via the concept of subsystem purity.
For multi-partite systems, evaluation of the purity of subsystems has been proposed as a simple quantitative lower bound for the extent of entanglement that can be generated from a mixed state \cite{PhysRevLett.87.050401}. Therefore, we will use this as an additional metric alongside classical shadows to quantify the measurement error in an n-qubit photonic state.

For a quantum system $\mathcal{S}$ comprised of $n$ qubits with the purity of a subset $\mathcal{A}$ consisting of $n'$ qubits, subsystem purity $P(\rho_{\mathcal{A}})$ is defined by
\begin{equation}
    P(\rho_A) = \mathrm{Tr}[\rho^2_{\mathcal{A}}],
\end{equation}
where $\rho_{\mathcal{A}} = \mathrm{Tr}_{\mathcal{S} \setminus \mathcal{A}}(\rho)$ is the reduced density matrix of subsystem $\mathcal{A}$. The purity $\mathrm{Tr}(\rho^2_{\mathcal{A}})$ of the subsystem can be calculated using the statistics of the output strings $\textbf{b}$ of the measurements \cite{brydges2019probing}:
\begin{equation} \label{eq:purity}
    \mathrm{Tr}(\rho^2_{\mathcal{A}}) = 2^{n'} \mathbb{E}_{\mathbf{i}_s}\left[\sum_{\mathbf{i}_s, \mathbf{b}_s, \mathbf{b}_s'} (-2)^{D[\mathbf{b}_s, \mathbf{b}_s']}\,  P_{(\mathbf{i}_s,\mathbf{b}_s)} \,P_{(\mathbf{i}_s,\mathbf{b_s'})}\right].
\end{equation}
Here $\mathbf{i}_s = [i_1, i_2, \dots, i_{n'}]$ represents the measurement basis, $\mathbf{b}_s = [b_1, b_2, \dots, b_{n'}] \in \{0, 1 \}^{n'}$  represents the bit string of the readout restricted to the subsystem. The expectation value $\mathbb{E}_{\mathbf{i}}$  is an average overall $K$ elements of the measurement basis. $D[\mathbf{b}_s, \mathbf{b'}_s]$ is the Hamming distance between the two bit strings $\mathbf{b}_s$ and $\mathbf{b'}_s$. The probability $P_{(\mathbf{i}_s,\mathbf{b}_s)}$ is given by $P_{(\mathbf{i}_s,\mathbf{b}_s)} = \mathrm{Tr}[\rho_{\mathcal{A}} \, E_{(i_1,b_1)} \otimes \cdots \otimes E_{(i_{n'},b_{n'})}]$. Eq. \ref{eq:purity} shows that the subsystem purity equals a weighted average over the Hamming distance between bitstring outcomes for pairs of measurements in the same basis $i$.

Fig. \ref{purity} shows the subsystem purity evaluated for a four-qubit $W$ state using $2\times10^4$ measurement outcomes. In Fig. \ref{purity}, the dashed black dots indicate the ideal true purity. The green dots represent the estimated purity derived from noisy measurement outcomes without applying the error mitigation strategy. The red dots correspond to the error mitigation strategy results. The left panel (a) demonstrates that the purity, when mitigated for errors, approaches that of the original W state. This suggests that the error mitigation method eliminates noise-induced bias in the purity estimation. The right panel (b) shows the robustness of the protocol across a wide range of error scale values, $h = |\boldsymbol{\lambda^\prime}|/|\boldsymbol{\lambda_{opt}}|$. These results indicate that our protocol enables an unbiased estimation of quantum state properties and is effective across a broad range of system sizes and noise levels.

Through this comprehensive analysis, we have demonstrated that our error mitigation protocol for the metasurface measurement of photonic states efficiently improves the estimation of quantum state properties in the presence of noise, displaying its applicability and robustness for diverse observables and a broad range of noisy environments. This exemplifies the protocol's potential for state property estimation for quantum computing and communication tasks.

\section{Discussion}\label{discussion}

The current work primarily focused on the intrinsic noise associated with metasurfaces. This inherent noise includes unwanted diffraction and photon loss, phenomena that persist even under flawless metasurface fabrication and are typically stochastic. On the other hand, fabrication imperfections tend to result in deviations from the desired measurement basis, which present as coherent rather than stochastic noise. It is important to note that our measurement bases are tomographically complete. This means that it is also, in principle, possible to calibrate the coherent measurement noise and model its impact on the measurement outcomes, thereby facilitating the mitigation of noise induced by fabrication imperfections.

Some randomized measurement protocols need nonlocal measurements. These typically involve state manipulation via a unitary operation capable of generating entanglement before measurements \cite{elben2023randomized}. However, metasurfaces can achieve multi-photon interference \cite{wang2018quantum}, a characteristic that enables direct nonlocal measurements. We expect that the scheme described in this paper can be generalized to protocols requiring nonlocal randomized measurements, such as Clifford measurement-based classical shadow of quantum states \cite{huang2020predicting}.

In this work, we have described all the procedures required to validate the proposed protocol for measuring arbitrary quantum states of light with current quantum photonic technologies. Experimental implementation with small numbers of photons appears readily implementable today. We expect that such near-term implementation will enable further refinements and expansions of the protocol.

To summarize, we have presented an efficient protocol to perform randomized measurements using metasurfaces, which enables the measurement and characterization of arbitrary quantum states of photonic qubits. We demonstrated the feasibility and scalability of the approach, which eliminates the need for optical setup reconfiguration, thereby overcoming a significant challenge to achieving experimental scalability. To counter the effects of measurement noise arising in implementation with realistic metasurfaces, we developed a protocol that first establishes an effective noise model from calibrated measurements. Then, we used this to develop a corresponding error mitigation technique that ensures the accuracy of measurement outcomes and high fidelity. Results of realistic simulations with this approach show that the metasurface-based randomized measurement protocol provides a promising strategy for robust and resource-efficient estimation of the properties of photonic quantum states.

\section{Acknowledgements}
This material is based upon work supported by the U.S. Department of Energy, Office of Science, National Quantum Information Science Research Centers, and Quantum Systems Accelerator (HR, YP, and KBW). 
ZZ acknowledges the support from the Royal Society scholarship. LX and MR acknowledge support from the UK Research and Innovation Future Leaders Fellowship (MR/T040513/1).
The authors thank Ying Li and Quanwei Li for their valuable discussions.

\appendix

\setcounter{table}{0}
\renewcommand{\thetable}{A\arabic{table}}
\setcounter{figure}{0}
\renewcommand{\thefigure}{A\arabic{figure}}

\section{Design and characterization of metasurface}\label{appendix_metasurface}

Nanodisks are structures made of phase-changing materials, and when nanodisks are arranged in a periodic structure, we refer to them as metagratings. It can impose a spatially varying phase shift on incoming light waves. The metagrating functions like a unique diffractive grating, capable of directing incoming photons in different directions based on their polarization.

Consider a metagrating designed to differentiate between horizontal (H) and vertical (V) polarizations. When a photon, represented as $|\psi_1\rangle$, interacts with this metagrating, it's projected in one of two directions. The probability of it moving in each direction is $|\langle \psi_1|H\rangle|^2$ and $|\langle \psi_1|V\rangle|^2$, respectively.

Our metasurface incorporates three types of gratings: $\{(H,V), (H+V, H-V), (LC, RC)\}$. These are evenly distributed on a two-dimensional surface. When a photon encounters the metasurface, it is equally likely to interact with any of the three gratings, thereby undergoing randomized measurements in three pairs of bases.

The guiding principle in our metasurface design is to create a metasurface that projects photons under randomly chosen basis pairs of H/V, H$\pm$V, LC/RC. Our design involves a metasurface composed of periodic Si nanodisks, the parameters of which are tuned to optimize measurement efficiency. We use numerical simulations to assist the metasurface design.

\begin{figure*}
    \centering
    \includegraphics[width=16 cm]{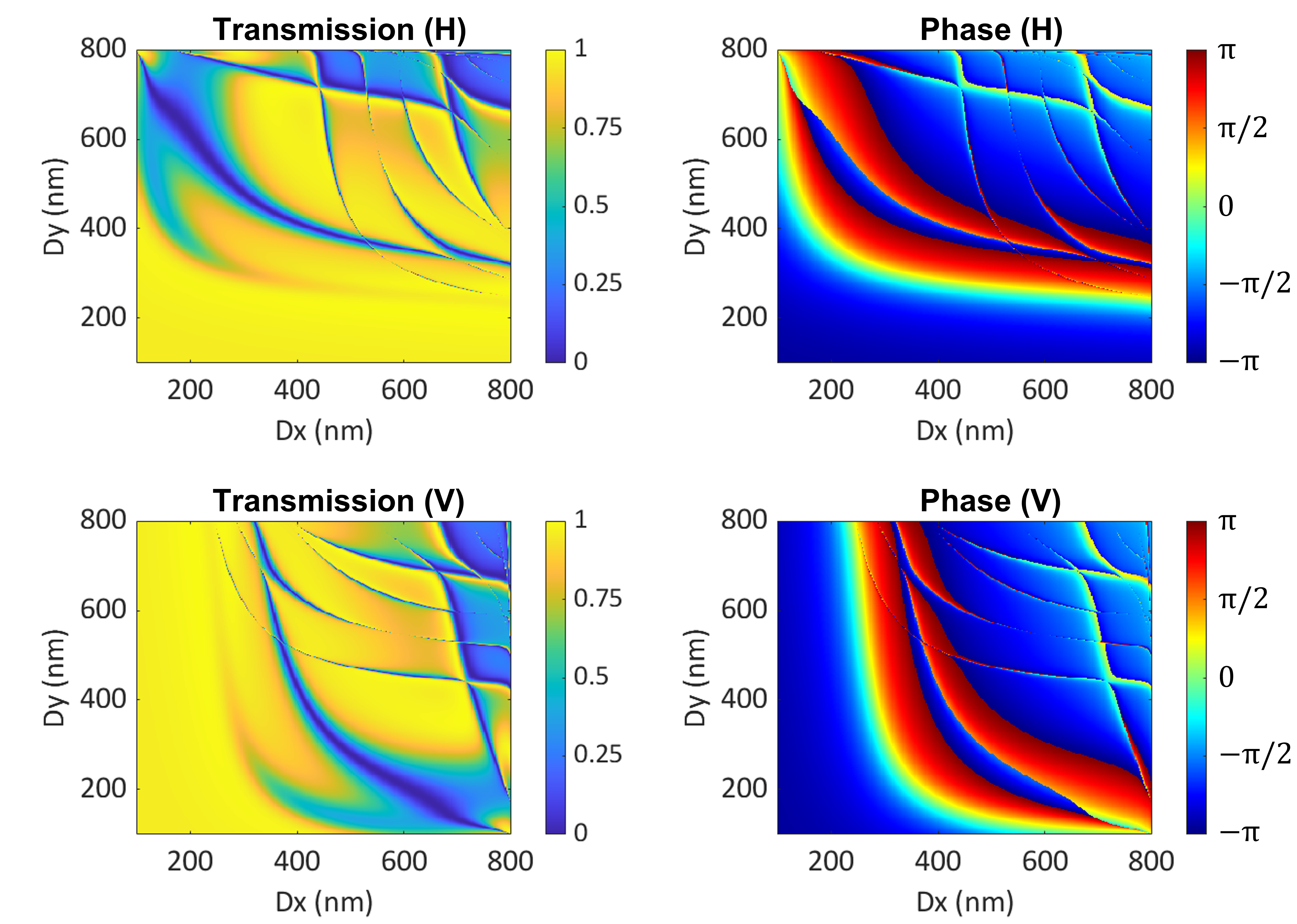}
    \caption{The transmission and phase change spectra of metasurfaces composed of nanodisks under x- (H) and y-polarised (V) plane wave, while tuning the width and depth of the nanodisks. The geometry of nanodisks is chosen to optimize the transmission and phase change.}
    \label{fig:Design}
\end{figure*}

For an incident wavelength of 1550nm, as an example, the nanodisks are designed to have a height $h_0=800$ nm and periodicity $p_0=800$ nm. We then scan the width and depth of the nanodisks along the x- ($D_x$) and y-axis ($D_y$), calculating the corresponding optical properties of the metasurfaces. Fig.  \ref{fig:Design} displays the transmission and phase change spectra under x- and y-polarised plane waves while altering the width and depth of the nanodisks. The design of nanodisks is optimized to enhance the high quality of both transmission and phase change. Due to the excitations of Mie-type electric and magnetic dipole resonances within the nanodisks and their interference, the backward scattering is strongly suppressed, leading to distinct phase changes along the x and y axes and forming a birefringent film with high transmission.

In order to separate photons with opposite polarization into +1 and -1 diffraction orders, we leverage the spatial linear dependence of the geometric phase on the transverse position of metagratings (along the x-axis). We utilize the genetic algorithm (AGA) to carefully select the geometric parameters of each unit to arrange a gradient phase distribution of two orthogonal states ($\phi_1$ and $\phi_2$) along the x-axis with high transmission. The unit size is 750 $\times$ 800 nm. We have designed metagratings that can separate three pairs of orthogonal states (HV, H$\pm$V, and RC/LC), illuminating into different angles with 4$\times$1, 5$\times$1, and 6$\times$1 meta-atoms as periodic units (as shown in Fig.  \ref{ms3}). By assembling these metagratings, we can form metasurfaces that distinguish three pairs of orthogonal states into different diffractive angles, leading to six ports for characterizing quantum states (as shown in Fig.  \ref{ms3}).

The detailed parameters of the metasurface design are listed in Table  \ref{design params}. Given the current experimental capabilities \cite{zheng2023advances}, it is feasible to fabricate a metasurface with these parameters. With this design, we numerically characterize the measurement efficiency, as shown in Table  \ref{characterization}. The data used for noise calibration are listed in Table  \ref{calibration data}, and the calibrated noise values can be found in Table  \ref{noise rate}.

The transmission and phase change spectra, obtained while adjusting the width and depth of the nanodisk, are calculated using Rigorous Coupled-Wave Analysis (RCWA) \cite{moharam1995formulation}. RCWA is a frequency-domain modal method based on the decomposition of a periodic structure and the pseudo-periodic solution of Maxwell's equations in terms of their Fourier expansions \cite{hugonin2021reticolo}.

The optical properties of metagratings are computed using the finite element method via COMSOL Multiphysics 6.0 software \cite{multiphysics1998introduction}.

\begin{centering}

\begin{table*}[]
\centering
\begin{tabular}{cccccccc}
\hline
\multicolumn{2}{c}{Grating }     & 1   & 2   & 3   & 4   & 5                 & 6                 \\ \hline
\multirow{3}{*}{HV}   & Dx (nm) & 570 & 253 & 340 & 328 & \multirow{3}{*}{} & \multirow{3}{*}{} \\
                      & Dy (nm) & 264 & 285 & 236 & 347 &                   &                   \\
                      & $\theta$     & 0   & 0   & 0   & 0   &                   &                   \\ \hline
\multirow{3}{*}{H±V}  & Dx (nm) & 257 & 320 & 341 & 111 & 397               & \multirow{3}{*}{} \\
                      & Dy (nm) & 375 & 314 & 266 & 400 & 102               &                   \\
                      & $\theta$     & 45  & 45  & 45  & 45  & 45                &                   \\ \hline
\multirow{3}{*}{RCLC} & Dx (nm) & 355 & 355 & 355 & 355 & 355               & 355               \\
                      & Dy (nm) & 253 & 253 & 253 & 253 & 253               & 253               \\
                      & $\theta$    & 0   & 30  & 60  & 90  & 120               & 150               \\ \hline
\end{tabular}
\caption{The geometric parameters of three metagratings. D$_x$ and D$_y$ are separately the width and depth of the nanodisk. $\theta$ is the rotation angle between the geometric and x axes.}
\label{design params}
\end{table*}
\end{centering}

\begin{table*}[]
\centering
\begin{tabular}{llllllll}
\hline
\multicolumn{2}{l}{Incident polarization} & H       & V       & H+V     & H-V     & RC      & LC      \\ \hline
\multirow{2}{*}{Grating HV}    & T{+1,0}  & 0.76556 & 0.01533 & 0.39063 & 0.39026 & 0.39053 & 0.39037 \\
                               & T{-1,0}  & 0.01058 & 0.84082 & 0.42521 & 0.42619 & 0.4256  & 0.4258  \\ \hline
\multirow{2}{*}{Grating H$\pm$V}       & T{+1,0}  & 0.36103 & 0.36608 & 0.69704 & 0.03006 & 0.39121 & 0.33588 \\
                               & T{-1,0}  & 0.40301 & 0.43212 & 0.05635 & 0.7788  & 0.40897 & 0.42618 \\ \hline
\multirow{2}{*}{Grating RCLC}  & T{+1,0}  & 0.38659 & 0.38188 & 0.39676 & 0.37172 & 0.00028 & 0.76819 \\
                               & T{-1,0}  & 0.3874  & 0.38048 & 0.37189 & 0.39599 & 0.7676  & 0.00028 \\ \hline 
\end{tabular}
\caption{The transmission of diffraction order {+1,0} and {-1,0} under the illumination of plane waves with different polarization states on three types of gratings. To illustrate, when
horizontally polarized photons hit the HV grating, 76.556 $\%$ of them are projected onto the dominant diffractive order
of one H side, with 1.058 $\%$ of them on the other V side. While there are multiple diffractive orders, we focus on the first two dominant ones ($\pm1$, 0), as higher orders have diminishing probabilities that are hard to detect and contribute to photon loss}
\label{characterization} 
\end{table*}

\begin{table*}[]
\centering
\begin{tabular}{ccccccc}
\hline
    & \multicolumn{6}{c}{Ouput photon count}        \\
\begin{tabular}[c]{@{}c@{}}Incident polarization\\ (input 10k photons)\end{tabular} & H & V & H+V & H-V & RC & LC \\ \hline
H   & 2552 & 35   & 1203 & 1343 & 1291 & 1289 \\
V   & 51   & 2803 & 1220 & 1440 & 1268 & 1273 \\
H+V & 1302 & 1417 & 2323 & 188  & 1240 & 1323 \\
H-V & 1301 & 1421 & 100  & 2596 & 1320 & 1239 \\
RC  & 1302 & 1419 & 1304 & 1363 & 2559 & 1    \\
LC  & 1301 & 1419 & 1120 & 1421 & 1    & 2561 \\ \hline
\end{tabular}
\caption{The calculated transmitted photon numbers on six ports with ten thousand incident photons collected under six bases (H, V, H+V, H-V, RC, LC). The numbers are calculated based on the simulated transmission of diffraction order {+1,0} and {-1,0} of three types of gratings in Table \ref{characterization}. For instance, if 10,000 horizontally polarized photons are incident, about one-third interact with the HV grating. The interaction directs 76.556 $\%$ photons to the H side and 1.058 $\%$ to the V side, producing the $(10000/3)*76.556\% \approx 2552$ on the left-up data of the table.}
\label{calibration data}
\end{table*}

\begin{table*}[]
\centering
\begin{tabular}{lllllll}
\hline
Noise         & H        & V        & H+V      & H-V      & RC       & LC       \\ \hline
Effective basis-flip & 0.012466 & 0.012466 & 0.054692 & 0.054692 & 0.000383 & 0.000383 \\
Amplitude damping  & 7.14E-03 & 7.14E-03 & 0        & 0        & 1.46E-05 & 1.46E-05 \\
Photon loss        & 0.223475 & 0.144225 & 0.275352 & 0.162548 & 0.234566 & 0.228934 \\ \hline
\end{tabular}
\caption{Calibrated noise values. The photon loss value, measured in units per injected photon, indicates the probability of photon loss}
\label{noise rate}
\end{table*}

\section{MEASUREMENT CHANNEL AND QUANTUM 2-DESIGN}\label{2design}

In this section, we demonstrate that the quantum channel corresponding to the measurements defined by the positive operator-valued measure (POVM) $\mathbf{E}$ behaves as a depolarizing channel when $\mathbf{E}$ forms a quantum 2-design. This result, while extensively examined through the lens of representation theory, is briefly derived here for the sake of comprehensiveness.

We begin by defining a quantum 2-design. Consider a set of vectors $\{|\phi_i\rangle \}_{i=1}^K$, residing within the unit sphere $\mathbb{S}^d$ of a $d$-dimensional Hilbert space $\mathcal{H}^d$. This set is said to form a quantum 2-design if it meets the following condition  \cite{scott2006tight}

\begin{equation}
    \frac{1}{K} \sum_{i=1}^{K}|\phi_i \rangle \langle \phi_i| ^{\otimes 2} = \int_{\mathbb{S}^d} |\phi \rangle \langle \phi| ^{\otimes 2} d\mu(\phi)\label{haar},
\end{equation}
where $d\mu(\phi)$ is the uniform spherical measure defined on $\mathbb{S}^d$ \cite{watrous2018theory}. A simple criterion for testing whether a set of states forms a quantum 2-design is given by the following equation \cite{gross2007evenly}
\begin{equation}
    \sum_{k, k^{\prime}}\left|\left\langle\phi_k \mid \phi_{k^{\prime}}\right\rangle\right|^4 / K^2=2 /\left(d^4+d^2\right).
\end{equation}

The Haar integral in Eq. \ref{haar} can be evaluated using results from representation theory  \cite{gross2015partial} to give
\begin{equation}
\begin{aligned}
    \int_{\mathbb{S}^d} |\phi \rangle \langle \phi| ^{\otimes 2} d\mu(\phi) &= \left(\begin{array}{c}
d+1 \\
2
\end{array}\right)^{-1}  \Pi_{\mathrm{Sym}^{(2)}}  \\
&= \left(\begin{array}{c}
d+1 \\
2
\end{array}\right)^{-1} \frac{I + W }{2},
\end{aligned}
\end{equation}
where $\Pi_{\mathrm{Sym}^{(2)}}$ is the projector onto the symmetric subspace of $\mathcal{H}^d \otimes \mathcal{H}^d$, and $W $ is the swap operator defined on $\mathcal{H}^d \otimes \mathcal{H}^d$, i.e. $W|\phi_i\rangle \otimes |\phi_j\rangle = |\phi_j\rangle \otimes |\phi_i\rangle, \forall |\phi_j\rangle , |\phi_i\rangle \in \mathcal{H}^d$.

For randomized measurements of a single qubit,  we can evaluate the measurement channel $\hat{M}$ defined by the quantum 2-design POVM $\mathbf{E}$ using the above results
\begin{equation}
\begin{aligned}
    \hat{M}(\rho) &= \frac{1}{K}
    \sum_{i = 1}^K \mathrm{Tr}\left(\rho |\phi_i \rangle \langle \phi_i | \right) |\phi_i \rangle \langle \phi_i | \\
    &= \mathrm{Tr}_1 \left[ (\rho \otimes I) \cdot \left(\frac{1}{K}
    \sum_{i = 1}^K  |\phi_i \rangle \langle \phi_i | ^{\otimes 2}\right) \right] \\
    &= \frac{I  +  \rho}{3},
\end{aligned}
\end{equation}
where $\mathrm{Tr}_1$ denotes the partial trace over the first copy of the system, and we take $d = 2$ for the single qubit case. This allows us to invert the measurement channel as $\hat{M}^{-1} (X) = 3X - I$, which is the same as in the unitary-based classical shadows, as expected. 

\begin{figure*}
\centering
\begin{subfigure}{.3\textwidth}
  \centering
  \includegraphics[width=5.5cm]{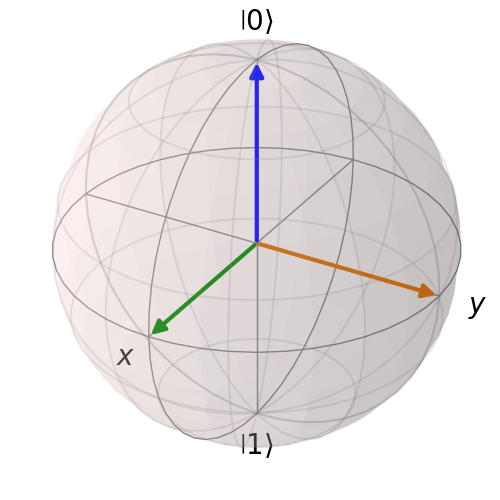}
  \caption{6 ports}
  \label{}
\end{subfigure}%
\begin{subfigure}{.3\textwidth}
  \centering
  \includegraphics[width=5.5cm]{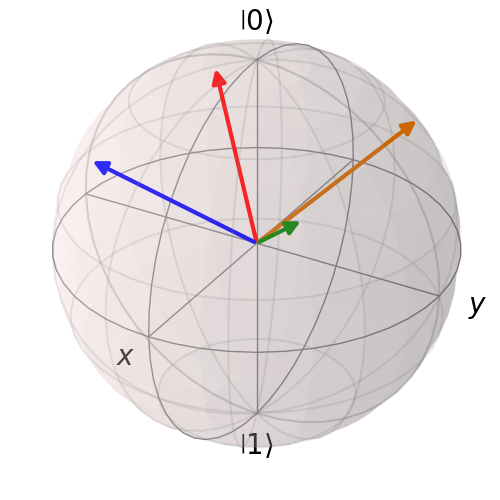}
  \caption{8 ports}
  \label{}
\end{subfigure}
\begin{subfigure}{.3\textwidth}
  \centering
  \includegraphics[width=5.5cm]{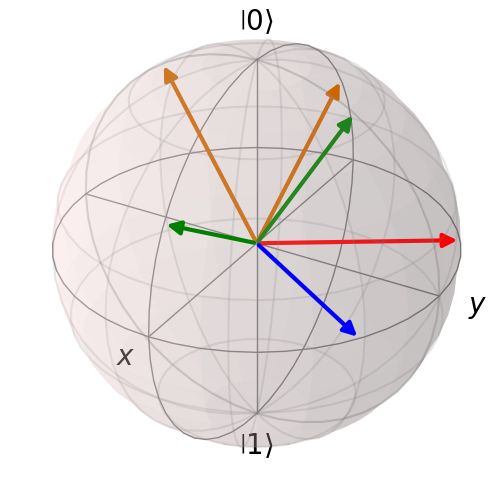}
  \caption{12 ports}
  \label{}
\end{subfigure}
\caption{Visulazation of the measurement bases of metasurface in the Bloch sphere. Vectors $|v\rangle$, corresponding to the POVM element $\frac{2}{K}|v\rangle \langle v|$, are represented in the Bloch sphere. Combined with their opposing vectors, which are omitted here, they constitute a complete POVM set. The arrangements of these vectors adhere to a spherical 2-design, thereby resulting in the formation of geometric configurations: Octahedron (6 ports), Hexahedron (8 ports), and Icosahedron (12 ports).}
\label{fig:spherical designs}
\end{figure*}

\begin{figure*}
\centering
\begin{subfigure}{.5\textwidth}
  \centering
  \includegraphics[width=8.6cm]{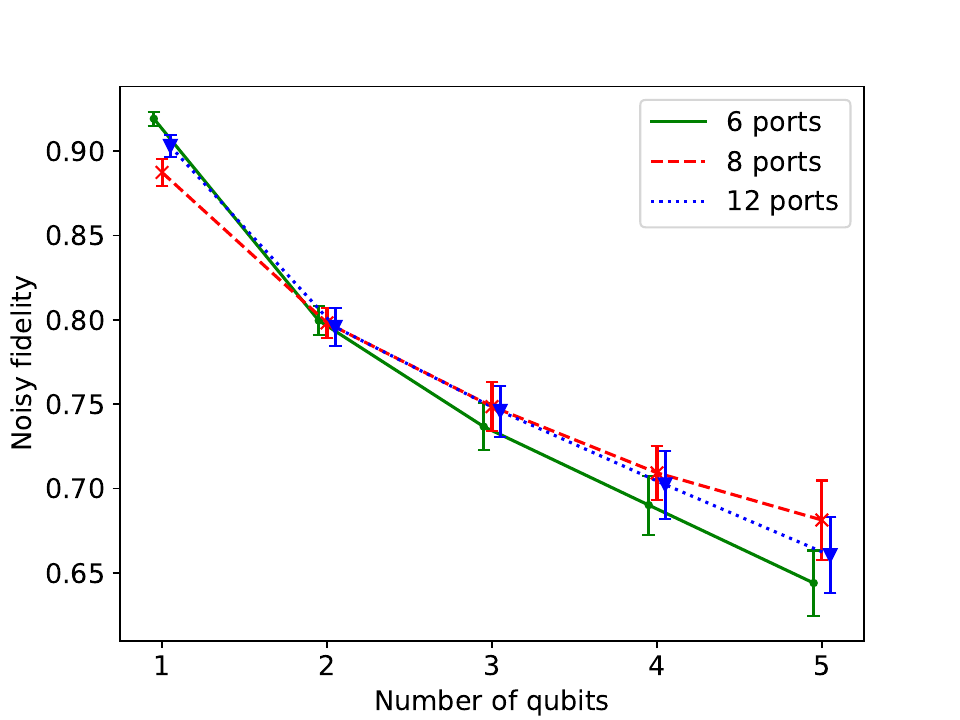}
  \caption{Fidelity estimation without error mitigation}
  \label{}
\end{subfigure}%
\begin{subfigure}{.5\textwidth}
  \centering
  \includegraphics[width=8.6cm]{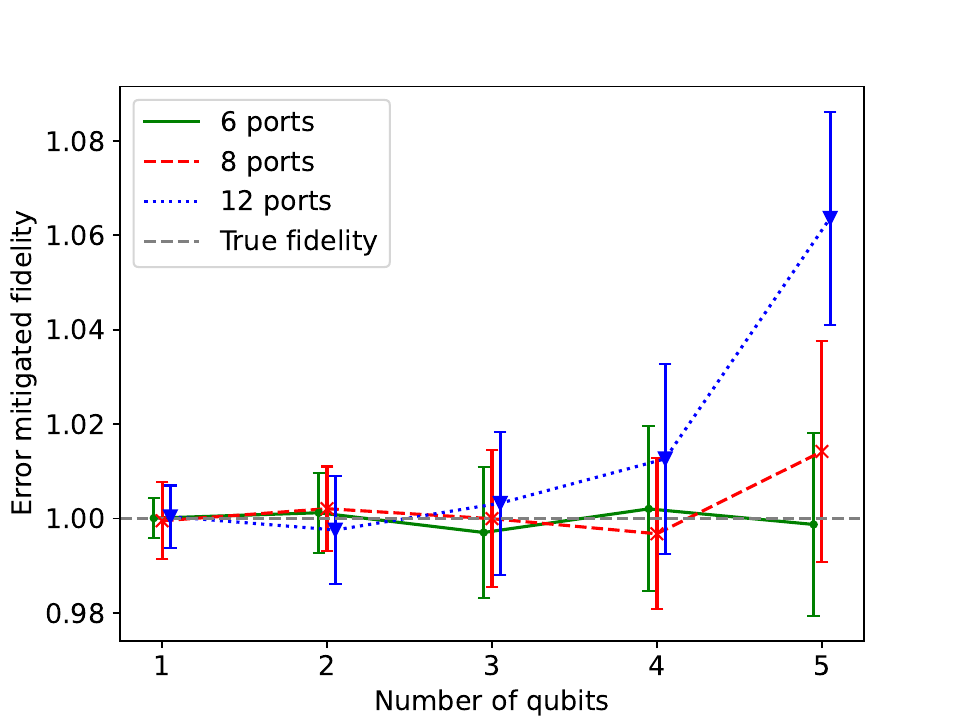}
  \caption{Fidelity estimation with error mitigation}
  \label{}
\end{subfigure}
\caption{Evaluation of W-State Fidelity through Randomized Measurements Using Various Metasurface Configurations. Metasurfaces, designed based on a spherical 2-design in the Stokes sphere, with different numbers of polarization measure bases (6, 8, and 12), produce valid randomized measurements due to their proven equivalence to a quantum 2-design [See Appendix \ref{2design}]. Error mitigation is feasible across all port numbers, but it is most effective with metasurfaces featuring six ports.}
\label{fig:fidelity-2design}
\end{figure*}

In the case of a single qubit, the quantum 2-designs can be mapped to spherical 2-designs on the Bloch/Stokes sphere \cite{foreman2015optimal, nguyen2022optimizing}. The projectors in the POVM for $K = 6$, $8$, $12$ are depicted as vectors on the Bloch sphere in Fig \ref{fig:spherical designs}. We have also conducted numerical simulations to estimate the fidelity of multi-qubit W-state, including noise, using POVMs with $K = 6$, $8$, $12$, with the results presented in Fig. \ref{fig:fidelity-2design}. The performance of the different POVMs is quite similar, although the 8-port POVM shows a slight advantage in the absence of error mitigation, and the 6-port POVM performs marginally better when error mitigation is employed. 

\begin{figure*}
\centering
\begin{subfigure}{.5\textwidth}
  \centering
  \includegraphics[width=8.6cm]{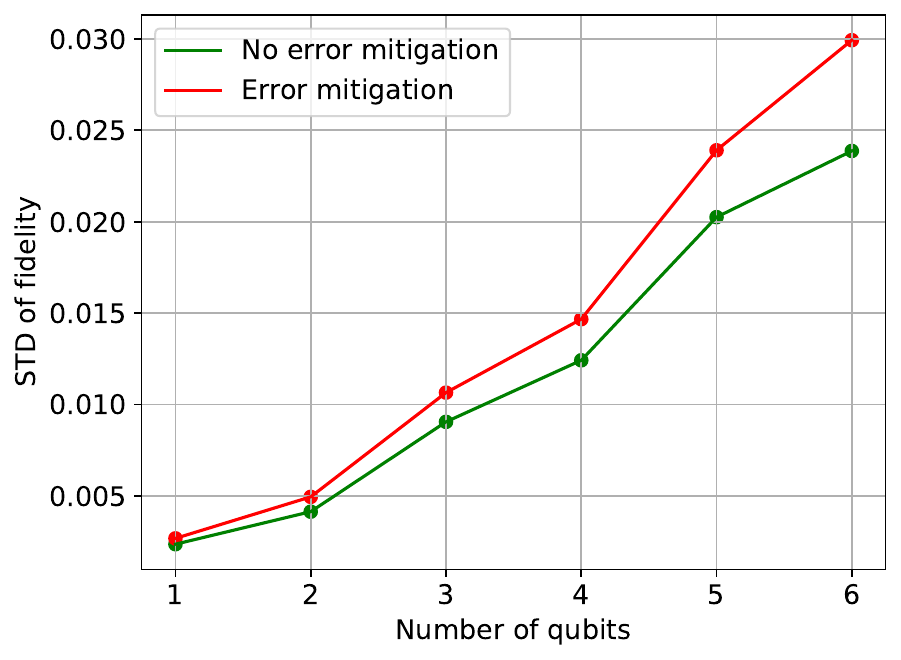}
  \caption{}
  \label{}
\end{subfigure}%
\begin{subfigure}{.5\textwidth}
  \centering
  \includegraphics[width=8.6cm]{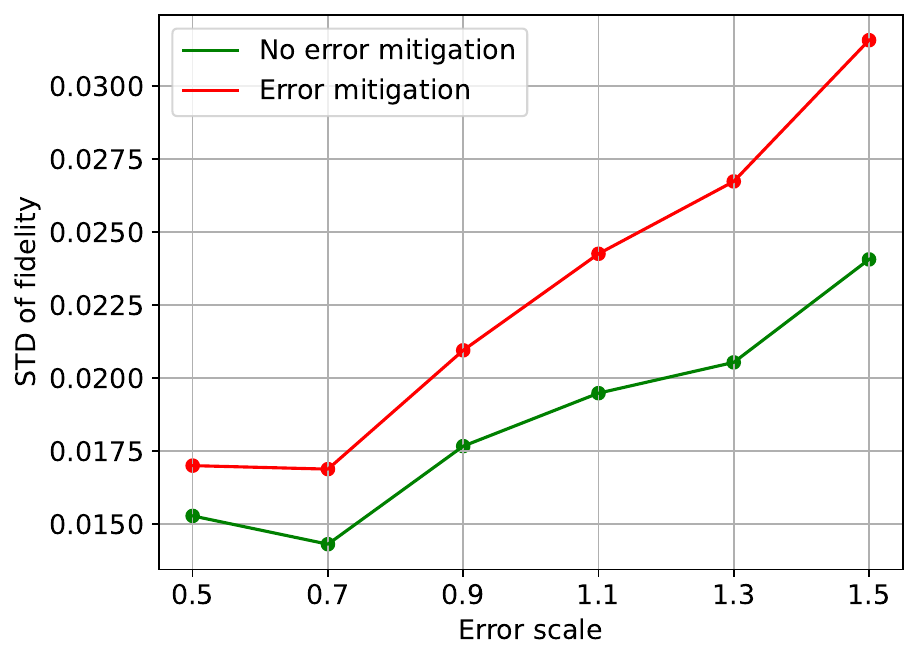}
  \caption{}
  \label{}
\end{subfigure}
\caption{Quantifying Sample Complexity in Error Mitigation Protocols. The implementation of error mitigation techniques effectively eliminates noise-induced bias, albeit at the cost of increasing the standard deviation (STD) of the estimation, thus highlighting an increase in sample complexity. The fidelity STD of error mitigation is approximately 1.5 times higher than those without, suggesting that the additional resources required for error mitigation implementation are justifiable.}
\label{fig:sample-complexity}
\end{figure*}

Fig. \ref{fig:fidelity-2design} demonstrates that across different metasurface designs, the protocol consistently and accurately estimates the properties of photonic states, highlighting its versatility and broad applicability. Figure \ref{fig:sample-complexity} illustrates that the incorporation of error mitigation incurs only a moderate increase in experimental resource costs.

\clearpage
\bibliography{main}%

\begin{thebibliography}{42}%
\makeatletter
\providecommand \@ifxundefined [1]{%
 \@ifx{#1\undefined}
}%
\providecommand \@ifnum [1]{%
 \ifnum #1\expandafter \@firstoftwo
 \else \expandafter \@secondoftwo
 \fi
}%
\providecommand \@ifx [1]{%
 \ifx #1\expandafter \@firstoftwo
 \else \expandafter \@secondoftwo
 \fi
}%
\providecommand \natexlab [1]{#1}%
\providecommand \enquote  [1]{``#1''}%
\providecommand \bibnamefont  [1]{#1}%
\providecommand \bibfnamefont [1]{#1}%
\providecommand \citenamefont [1]{#1}%
\providecommand \href@noop [0]{\@secondoftwo}%
\providecommand \href [0]{\begingroup \@sanitize@url \@href}%
\providecommand \@href[1]{\@@startlink{#1}\@@href}%
\providecommand \@@href[1]{\endgroup#1\@@endlink}%
\providecommand \@sanitize@url [0]{\catcode `\\12\catcode `\$12\catcode
  `\&12\catcode `\#12\catcode `\^12\catcode `\_12\catcode `\%12\relax}%
\providecommand \@@startlink[1]{}%
\providecommand \@@endlink[0]{}%
\providecommand \url  [0]{\begingroup\@sanitize@url \@url }%
\providecommand \@url [1]{\endgroup\@href {#1}{\urlprefix }}%
\providecommand \urlprefix  [0]{URL }%
\providecommand \Eprint [0]{\href }%
\providecommand \doibase [0]{https://doi.org/}%
\providecommand \selectlanguage [0]{\@gobble}%
\providecommand \bibinfo  [0]{\@secondoftwo}%
\providecommand \bibfield  [0]{\@secondoftwo}%
\providecommand \translation [1]{[#1]}%
\providecommand \BibitemOpen [0]{}%
\providecommand \bibitemStop [0]{}%
\providecommand \bibitemNoStop [0]{.\EOS\space}%
\providecommand \EOS [0]{\spacefactor3000\relax}%
\providecommand \BibitemShut  [1]{\csname bibitem#1\endcsname}%
\let\auto@bib@innerbib\@empty
\bibitem [{\citenamefont {Feynman}(1982)}]{feyman}%
  \BibitemOpen
  \bibfield  {author} {\bibinfo {author} {\bibfnamefont {R.~P.}\ \bibnamefont
  {Feynman}},\ }\bibfield  {title} {\bibinfo {title} {Simulating physics with
  computers},\ }\href {https://doi.org/10.1007/BF02650179} {\bibfield
  {journal} {\bibinfo  {journal} {International Journal of Theoretical Physics
  volume}\ }\textbf {\bibinfo {volume} {227}},\ \bibinfo {pages} {109}
  (\bibinfo {year} {1982})}\BibitemShut {NoStop}%
\bibitem [{\citenamefont {Shor}(1994)}]{365700}%
  \BibitemOpen
  \bibfield  {author} {\bibinfo {author} {\bibfnamefont {P.}~\bibnamefont
  {Shor}},\ }\bibfield  {title} {\bibinfo {title} {Algorithms for quantum
  computation: discrete logarithms and factoring},\ }in\ \href
  {https://doi.org/10.1109/SFCS.1994.365700} {\emph {\bibinfo {booktitle}
  {Proceedings 35th Annual Symposium on Foundations of Computer Science}}}\
  (\bibinfo {year} {1994})\ pp.\ \bibinfo {pages} {124--134}\BibitemShut
  {NoStop}%
\bibitem [{\citenamefont {Elben}\ \emph {et~al.}(2023)\citenamefont {Elben},
  \citenamefont {Flammia}, \citenamefont {Huang}, \citenamefont {Kueng},
  \citenamefont {Preskill}, \citenamefont {Vermersch},\ and\ \citenamefont
  {Zoller}}]{elben2023randomized}%
  \BibitemOpen
  \bibfield  {author} {\bibinfo {author} {\bibfnamefont {A.}~\bibnamefont
  {Elben}}, \bibinfo {author} {\bibfnamefont {S.~T.}\ \bibnamefont {Flammia}},
  \bibinfo {author} {\bibfnamefont {H.-Y.}\ \bibnamefont {Huang}}, \bibinfo
  {author} {\bibfnamefont {R.}~\bibnamefont {Kueng}}, \bibinfo {author}
  {\bibfnamefont {J.}~\bibnamefont {Preskill}}, \bibinfo {author}
  {\bibfnamefont {B.}~\bibnamefont {Vermersch}},\ and\ \bibinfo {author}
  {\bibfnamefont {P.}~\bibnamefont {Zoller}},\ }\bibfield  {title} {\bibinfo
  {title} {The randomized measurement toolbox},\ }\href@noop {} {\bibfield
  {journal} {\bibinfo  {journal} {Nature Reviews Physics}\ }\textbf {\bibinfo
  {volume} {5}},\ \bibinfo {pages} {9} (\bibinfo {year} {2023})}\BibitemShut
  {NoStop}%
\bibitem [{\citenamefont {Huang}\ \emph {et~al.}(2020)\citenamefont {Huang},
  \citenamefont {Kueng},\ and\ \citenamefont {Preskill}}]{huang2020predicting}%
  \BibitemOpen
  \bibfield  {author} {\bibinfo {author} {\bibfnamefont {H.-Y.}\ \bibnamefont
  {Huang}}, \bibinfo {author} {\bibfnamefont {R.}~\bibnamefont {Kueng}},\ and\
  \bibinfo {author} {\bibfnamefont {J.}~\bibnamefont {Preskill}},\ }\bibfield
  {title} {\bibinfo {title} {Predicting many properties of a quantum system
  from very few measurements},\ }\href@noop {} {\bibfield  {journal} {\bibinfo
  {journal} {Nature Physics}\ }\textbf {\bibinfo {volume} {16}},\ \bibinfo
  {pages} {1050} (\bibinfo {year} {2020})}\BibitemShut {NoStop}%
\bibitem [{\citenamefont {Zhang}\ \emph {et~al.}(2021)\citenamefont {Zhang},
  \citenamefont {Sun}, \citenamefont {Fang}, \citenamefont {Zhang},
  \citenamefont {Yuan},\ and\ \citenamefont {Lu}}]{PhysRevLett.127.200501}%
  \BibitemOpen
  \bibfield  {author} {\bibinfo {author} {\bibfnamefont {T.}~\bibnamefont
  {Zhang}}, \bibinfo {author} {\bibfnamefont {J.}~\bibnamefont {Sun}}, \bibinfo
  {author} {\bibfnamefont {X.-X.}\ \bibnamefont {Fang}}, \bibinfo {author}
  {\bibfnamefont {X.-M.}\ \bibnamefont {Zhang}}, \bibinfo {author}
  {\bibfnamefont {X.}~\bibnamefont {Yuan}},\ and\ \bibinfo {author}
  {\bibfnamefont {H.}~\bibnamefont {Lu}},\ }\bibfield  {title} {\bibinfo
  {title} {Experimental quantum state measurement with classical shadows},\
  }\href {https://doi.org/10.1103/PhysRevLett.127.200501} {\bibfield  {journal}
  {\bibinfo  {journal} {Phys. Rev. Lett.}\ }\textbf {\bibinfo {volume} {127}},\
  \bibinfo {pages} {200501} (\bibinfo {year} {2021})}\BibitemShut {NoStop}%
\bibitem [{\citenamefont {Struchalin}\ \emph {et~al.}(2021)\citenamefont
  {Struchalin}, \citenamefont {Zagorovskii}, \citenamefont {Kovlakov},
  \citenamefont {Straupe},\ and\ \citenamefont {Kulik}}]{PRXQuantum.2.010307}%
  \BibitemOpen
  \bibfield  {author} {\bibinfo {author} {\bibfnamefont {G.}~\bibnamefont
  {Struchalin}}, \bibinfo {author} {\bibfnamefont {Y.~A.}\ \bibnamefont
  {Zagorovskii}}, \bibinfo {author} {\bibfnamefont {E.}~\bibnamefont
  {Kovlakov}}, \bibinfo {author} {\bibfnamefont {S.}~\bibnamefont {Straupe}},\
  and\ \bibinfo {author} {\bibfnamefont {S.}~\bibnamefont {Kulik}},\ }\bibfield
   {title} {\bibinfo {title} {Experimental estimation of quantum state
  properties from classical shadows},\ }\href
  {https://doi.org/10.1103/PRXQuantum.2.010307} {\bibfield  {journal} {\bibinfo
   {journal} {PRX Quantum}\ }\textbf {\bibinfo {volume} {2}},\ \bibinfo {pages}
  {010307} (\bibinfo {year} {2021})}\BibitemShut {NoStop}%
\bibitem [{\citenamefont {Wang}\ \emph {et~al.}(2018)\citenamefont {Wang},
  \citenamefont {Titchener}, \citenamefont {Kruk}, \citenamefont {Xu},
  \citenamefont {Chung}, \citenamefont {Parry}, \citenamefont {Kravchenko},
  \citenamefont {Chen}, \citenamefont {Solntsev}, \citenamefont {Kivshar} \emph
  {et~al.}}]{wang2018quantum}%
  \BibitemOpen
  \bibfield  {author} {\bibinfo {author} {\bibfnamefont {K.}~\bibnamefont
  {Wang}}, \bibinfo {author} {\bibfnamefont {J.~G.}\ \bibnamefont {Titchener}},
  \bibinfo {author} {\bibfnamefont {S.~S.}\ \bibnamefont {Kruk}}, \bibinfo
  {author} {\bibfnamefont {L.}~\bibnamefont {Xu}}, \bibinfo {author}
  {\bibfnamefont {H.-P.}\ \bibnamefont {Chung}}, \bibinfo {author}
  {\bibfnamefont {M.}~\bibnamefont {Parry}}, \bibinfo {author} {\bibfnamefont
  {I.~I.}\ \bibnamefont {Kravchenko}}, \bibinfo {author} {\bibfnamefont
  {Y.-H.}\ \bibnamefont {Chen}}, \bibinfo {author} {\bibfnamefont {A.~S.}\
  \bibnamefont {Solntsev}}, \bibinfo {author} {\bibfnamefont {Y.~S.}\
  \bibnamefont {Kivshar}}, \emph {et~al.},\ }\bibfield  {title} {\bibinfo
  {title} {Quantum metasurface for multiphoton interference and state
  reconstruction},\ }\href@noop {} {\bibfield  {journal} {\bibinfo  {journal}
  {Science}\ }\textbf {\bibinfo {volume} {361}},\ \bibinfo {pages} {1104}
  (\bibinfo {year} {2018})}\BibitemShut {NoStop}%
\bibitem [{\citenamefont {Zheng}\ \emph {et~al.}(2023)\citenamefont {Zheng},
  \citenamefont {Rocco}, \citenamefont {Ren}, \citenamefont {Sergaeva},
  \citenamefont {Zhang}, \citenamefont {Whaley}, \citenamefont {Ying},
  \citenamefont {de~Ceglia}, \citenamefont {De-Angelis}, \citenamefont
  {Rahmani} \emph {et~al.}}]{zheng2023advances}%
  \BibitemOpen
  \bibfield  {author} {\bibinfo {author} {\bibfnamefont {Z.}~\bibnamefont
  {Zheng}}, \bibinfo {author} {\bibfnamefont {D.}~\bibnamefont {Rocco}},
  \bibinfo {author} {\bibfnamefont {H.}~\bibnamefont {Ren}}, \bibinfo {author}
  {\bibfnamefont {O.}~\bibnamefont {Sergaeva}}, \bibinfo {author}
  {\bibfnamefont {Y.}~\bibnamefont {Zhang}}, \bibinfo {author} {\bibfnamefont
  {K.~B.}\ \bibnamefont {Whaley}}, \bibinfo {author} {\bibfnamefont
  {C.}~\bibnamefont {Ying}}, \bibinfo {author} {\bibfnamefont {D.}~\bibnamefont
  {de~Ceglia}}, \bibinfo {author} {\bibfnamefont {C.}~\bibnamefont
  {De-Angelis}}, \bibinfo {author} {\bibfnamefont {M.}~\bibnamefont {Rahmani}},
  \emph {et~al.},\ }\bibfield  {title} {\bibinfo {title} {Advances in nonlinear
  metasurfaces for imaging, quantum, and sensing applications},\ }\href@noop {}
  {\bibfield  {journal} {\bibinfo  {journal} {Nanophotonics}\ } (\bibinfo
  {year} {2023})}\BibitemShut {NoStop}%
\bibitem [{\citenamefont {Chen}\ \emph {et~al.}(2016)\citenamefont {Chen},
  \citenamefont {Taylor},\ and\ \citenamefont {Yu}}]{chen2016review}%
  \BibitemOpen
  \bibfield  {author} {\bibinfo {author} {\bibfnamefont {H.-T.}\ \bibnamefont
  {Chen}}, \bibinfo {author} {\bibfnamefont {A.~J.}\ \bibnamefont {Taylor}},\
  and\ \bibinfo {author} {\bibfnamefont {N.}~\bibnamefont {Yu}},\ }\bibfield
  {title} {\bibinfo {title} {A review of metasurfaces: physics and
  applications},\ }\href@noop {} {\bibfield  {journal} {\bibinfo  {journal}
  {Reports on progress in physics}\ }\textbf {\bibinfo {volume} {79}},\
  \bibinfo {pages} {076401} (\bibinfo {year} {2016})}\BibitemShut {NoStop}%
\bibitem [{\citenamefont {Wyderka}\ \emph {et~al.}(2023)\citenamefont
  {Wyderka}, \citenamefont {Ketterer}, \citenamefont {Imai}, \citenamefont
  {B{\"o}nsel}, \citenamefont {Jones}, \citenamefont {Kirby}, \citenamefont
  {Yu},\ and\ \citenamefont {G{\"u}hne}}]{wyderka2023complete}%
  \BibitemOpen
  \bibfield  {author} {\bibinfo {author} {\bibfnamefont {N.}~\bibnamefont
  {Wyderka}}, \bibinfo {author} {\bibfnamefont {A.}~\bibnamefont {Ketterer}},
  \bibinfo {author} {\bibfnamefont {S.}~\bibnamefont {Imai}}, \bibinfo {author}
  {\bibfnamefont {J.~L.}\ \bibnamefont {B{\"o}nsel}}, \bibinfo {author}
  {\bibfnamefont {D.~E.}\ \bibnamefont {Jones}}, \bibinfo {author}
  {\bibfnamefont {B.~T.}\ \bibnamefont {Kirby}}, \bibinfo {author}
  {\bibfnamefont {X.-D.}\ \bibnamefont {Yu}},\ and\ \bibinfo {author}
  {\bibfnamefont {O.}~\bibnamefont {G{\"u}hne}},\ }\bibfield  {title} {\bibinfo
  {title} {Complete characterization of quantum correlations by randomized
  measurements},\ }\href@noop {} {\bibfield  {journal} {\bibinfo  {journal}
  {Physical Review Letters}\ }\textbf {\bibinfo {volume} {131}},\ \bibinfo
  {pages} {090201} (\bibinfo {year} {2023})}\BibitemShut {NoStop}%
\bibitem [{\citenamefont {Zhao}\ \emph {et~al.}(2022)\citenamefont {Zhao},
  \citenamefont {Geng}, \citenamefont {Wei}, \citenamefont {Liu}, \citenamefont
  {Zhou}, \citenamefont {Zhang}, \citenamefont {He}, \citenamefont {Li},
  \citenamefont {Li}, \citenamefont {Wang} \emph
  {et~al.}}]{zhao2022controllable}%
  \BibitemOpen
  \bibfield  {author} {\bibinfo {author} {\bibfnamefont {R.}~\bibnamefont
  {Zhao}}, \bibinfo {author} {\bibfnamefont {G.}~\bibnamefont {Geng}}, \bibinfo
  {author} {\bibfnamefont {Q.}~\bibnamefont {Wei}}, \bibinfo {author}
  {\bibfnamefont {Y.}~\bibnamefont {Liu}}, \bibinfo {author} {\bibfnamefont
  {H.}~\bibnamefont {Zhou}}, \bibinfo {author} {\bibfnamefont {X.}~\bibnamefont
  {Zhang}}, \bibinfo {author} {\bibfnamefont {C.}~\bibnamefont {He}}, \bibinfo
  {author} {\bibfnamefont {X.}~\bibnamefont {Li}}, \bibinfo {author}
  {\bibfnamefont {X.}~\bibnamefont {Li}}, \bibinfo {author} {\bibfnamefont
  {Y.}~\bibnamefont {Wang}}, \emph {et~al.},\ }\bibfield  {title} {\bibinfo
  {title} {Controllable polarization and diffraction modulated
  multi-functionality based on metasurface},\ }\href@noop {} {\bibfield
  {journal} {\bibinfo  {journal} {Advanced Optical Materials}\ }\textbf
  {\bibinfo {volume} {10}},\ \bibinfo {pages} {2102596} (\bibinfo {year}
  {2022})}\BibitemShut {NoStop}%
\bibitem [{\citenamefont {Nielsen}\ and\ \citenamefont
  {Chuang}(2010)}]{nielsen2010quantum}%
  \BibitemOpen
  \bibfield  {author} {\bibinfo {author} {\bibfnamefont {M.~A.}\ \bibnamefont
  {Nielsen}}\ and\ \bibinfo {author} {\bibfnamefont {I.~L.}\ \bibnamefont
  {Chuang}},\ }\href@noop {} {\emph {\bibinfo {title} {Quantum computation and
  quantum information}}}\ (\bibinfo  {publisher} {Cambridge university press},\
  \bibinfo {year} {2010})\BibitemShut {NoStop}%
\bibitem [{\citenamefont {Gross}\ \emph {et~al.}(2007)\citenamefont {Gross},
  \citenamefont {Audenaert},\ and\ \citenamefont {Eisert}}]{gross2007evenly}%
  \BibitemOpen
  \bibfield  {author} {\bibinfo {author} {\bibfnamefont {D.}~\bibnamefont
  {Gross}}, \bibinfo {author} {\bibfnamefont {K.}~\bibnamefont {Audenaert}},\
  and\ \bibinfo {author} {\bibfnamefont {J.}~\bibnamefont {Eisert}},\
  }\bibfield  {title} {\bibinfo {title} {Evenly distributed unitaries: On the
  structure of unitary designs},\ }\href@noop {} {\bibfield  {journal}
  {\bibinfo  {journal} {Journal of mathematical physics}\ }\textbf {\bibinfo
  {volume} {48}} (\bibinfo {year} {2007})}\BibitemShut {NoStop}%
\bibitem [{\citenamefont {Nguyen}\ \emph {et~al.}(2022)\citenamefont {Nguyen},
  \citenamefont {B{\"o}nsel}, \citenamefont {Steinberg},\ and\ \citenamefont
  {G{\"u}hne}}]{nguyen2022optimizing}%
  \BibitemOpen
  \bibfield  {author} {\bibinfo {author} {\bibfnamefont {H.~C.}\ \bibnamefont
  {Nguyen}}, \bibinfo {author} {\bibfnamefont {J.~L.}\ \bibnamefont
  {B{\"o}nsel}}, \bibinfo {author} {\bibfnamefont {J.}~\bibnamefont
  {Steinberg}},\ and\ \bibinfo {author} {\bibfnamefont {O.}~\bibnamefont
  {G{\"u}hne}},\ }\bibfield  {title} {\bibinfo {title} {Optimizing shadow
  tomography with generalized measurements},\ }\href@noop {} {\bibfield
  {journal} {\bibinfo  {journal} {Physical Review Letters}\ }\textbf {\bibinfo
  {volume} {129}},\ \bibinfo {pages} {220502} (\bibinfo {year}
  {2022})}\BibitemShut {NoStop}%
\bibitem [{\citenamefont {Elben}\ \emph {et~al.}(2019)\citenamefont {Elben},
  \citenamefont {Vermersch}, \citenamefont {Roos},\ and\ \citenamefont
  {Zoller}}]{elben2019statistical}%
  \BibitemOpen
  \bibfield  {author} {\bibinfo {author} {\bibfnamefont {A.}~\bibnamefont
  {Elben}}, \bibinfo {author} {\bibfnamefont {B.}~\bibnamefont {Vermersch}},
  \bibinfo {author} {\bibfnamefont {C.~F.}\ \bibnamefont {Roos}},\ and\
  \bibinfo {author} {\bibfnamefont {P.}~\bibnamefont {Zoller}},\ }\bibfield
  {title} {\bibinfo {title} {Statistical correlations between locally
  randomized measurements: A toolbox for probing entanglement in many-body
  quantum states},\ }\href@noop {} {\bibfield  {journal} {\bibinfo  {journal}
  {Physical Review A}\ }\textbf {\bibinfo {volume} {99}},\ \bibinfo {pages}
  {052323} (\bibinfo {year} {2019})}\BibitemShut {NoStop}%
\bibitem [{\citenamefont {Brydges}\ \emph {et~al.}(2019)\citenamefont
  {Brydges}, \citenamefont {Elben}, \citenamefont {Jurcevic}, \citenamefont
  {Vermersch}, \citenamefont {Maier}, \citenamefont {Lanyon}, \citenamefont
  {Zoller}, \citenamefont {Blatt},\ and\ \citenamefont
  {Roos}}]{brydges2019probing}%
  \BibitemOpen
  \bibfield  {author} {\bibinfo {author} {\bibfnamefont {T.}~\bibnamefont
  {Brydges}}, \bibinfo {author} {\bibfnamefont {A.}~\bibnamefont {Elben}},
  \bibinfo {author} {\bibfnamefont {P.}~\bibnamefont {Jurcevic}}, \bibinfo
  {author} {\bibfnamefont {B.}~\bibnamefont {Vermersch}}, \bibinfo {author}
  {\bibfnamefont {C.}~\bibnamefont {Maier}}, \bibinfo {author} {\bibfnamefont
  {B.~P.}\ \bibnamefont {Lanyon}}, \bibinfo {author} {\bibfnamefont
  {P.}~\bibnamefont {Zoller}}, \bibinfo {author} {\bibfnamefont
  {R.}~\bibnamefont {Blatt}},\ and\ \bibinfo {author} {\bibfnamefont {C.~F.}\
  \bibnamefont {Roos}},\ }\bibfield  {title} {\bibinfo {title} {Probing
  r{\'e}nyi entanglement entropy via randomized measurements},\ }\href@noop {}
  {\bibfield  {journal} {\bibinfo  {journal} {Science}\ }\textbf {\bibinfo
  {volume} {364}},\ \bibinfo {pages} {260} (\bibinfo {year}
  {2019})}\BibitemShut {NoStop}%
\bibitem [{\citenamefont {Maciejewski}\ \emph {et~al.}(2020)\citenamefont
  {Maciejewski}, \citenamefont {Zimbor{\'a}s},\ and\ \citenamefont
  {Oszmaniec}}]{maciejewski2020mitigation}%
  \BibitemOpen
  \bibfield  {author} {\bibinfo {author} {\bibfnamefont {F.~B.}\ \bibnamefont
  {Maciejewski}}, \bibinfo {author} {\bibfnamefont {Z.}~\bibnamefont
  {Zimbor{\'a}s}},\ and\ \bibinfo {author} {\bibfnamefont {M.}~\bibnamefont
  {Oszmaniec}},\ }\bibfield  {title} {\bibinfo {title} {Mitigation of readout
  noise in near-term quantum devices by classical post-processing based on
  detector tomography},\ }\href@noop {} {\bibfield  {journal} {\bibinfo
  {journal} {Quantum}\ }\textbf {\bibinfo {volume} {4}},\ \bibinfo {pages}
  {257} (\bibinfo {year} {2020})}\BibitemShut {NoStop}%
\bibitem [{\citenamefont {Geller}(2020)}]{geller2020rigorous}%
  \BibitemOpen
  \bibfield  {author} {\bibinfo {author} {\bibfnamefont {M.~R.}\ \bibnamefont
  {Geller}},\ }\bibfield  {title} {\bibinfo {title} {Rigorous measurement error
  correction},\ }\href@noop {} {\bibfield  {journal} {\bibinfo  {journal}
  {Quantum Science and Technology}\ }\textbf {\bibinfo {volume} {5}},\ \bibinfo
  {pages} {03LT01} (\bibinfo {year} {2020})}\BibitemShut {NoStop}%
\bibitem [{\citenamefont {Huang}\ \emph {et~al.}(2021)\citenamefont {Huang},
  \citenamefont {Xiao}, \citenamefont {Xie}, \citenamefont {Zheng},
  \citenamefont {Su}, \citenamefont {Chen}, \citenamefont {Liu}, \citenamefont
  {Tang},\ and\ \citenamefont {Li}}]{ma14092212}%
  \BibitemOpen
  \bibfield  {author} {\bibinfo {author} {\bibfnamefont {Y.}~\bibnamefont
  {Huang}}, \bibinfo {author} {\bibfnamefont {T.}~\bibnamefont {Xiao}},
  \bibinfo {author} {\bibfnamefont {Z.}~\bibnamefont {Xie}}, \bibinfo {author}
  {\bibfnamefont {J.}~\bibnamefont {Zheng}}, \bibinfo {author} {\bibfnamefont
  {Y.}~\bibnamefont {Su}}, \bibinfo {author} {\bibfnamefont {W.}~\bibnamefont
  {Chen}}, \bibinfo {author} {\bibfnamefont {K.}~\bibnamefont {Liu}}, \bibinfo
  {author} {\bibfnamefont {M.}~\bibnamefont {Tang}},\ and\ \bibinfo {author}
  {\bibfnamefont {L.}~\bibnamefont {Li}},\ }\bibfield  {title} {\bibinfo
  {title} {Reconfigurable continuous meta-grating for broadband polarization
  conversion and perfect absorption},\ }\bibfield  {journal} {\bibinfo
  {journal} {Materials}\ }\textbf {\bibinfo {volume} {14}},\ \href
  {https://doi.org/10.3390/ma14092212} {10.3390/ma14092212} (\bibinfo {year}
  {2021})\BibitemShut {NoStop}%
\bibitem [{\citenamefont {Smith}\ \emph {et~al.}(2021)\citenamefont {Smith},
  \citenamefont {Khosla}, \citenamefont {Self},\ and\ \citenamefont
  {Kim}}]{smith2021qubit}%
  \BibitemOpen
  \bibfield  {author} {\bibinfo {author} {\bibfnamefont {A.~W.}\ \bibnamefont
  {Smith}}, \bibinfo {author} {\bibfnamefont {K.~E.}\ \bibnamefont {Khosla}},
  \bibinfo {author} {\bibfnamefont {C.~N.}\ \bibnamefont {Self}},\ and\
  \bibinfo {author} {\bibfnamefont {M.}~\bibnamefont {Kim}},\ }\bibfield
  {title} {\bibinfo {title} {Qubit readout error mitigation with bit-flip
  averaging},\ }\href@noop {} {\bibfield  {journal} {\bibinfo  {journal}
  {Science advances}\ }\textbf {\bibinfo {volume} {7}},\ \bibinfo {pages}
  {eabi8009} (\bibinfo {year} {2021})}\BibitemShut {NoStop}%
\bibitem [{\citenamefont {Oreshkov}\ and\ \citenamefont
  {Brun}(2005)}]{oreshkov2005weak}%
  \BibitemOpen
  \bibfield  {author} {\bibinfo {author} {\bibfnamefont {O.}~\bibnamefont
  {Oreshkov}}\ and\ \bibinfo {author} {\bibfnamefont {T.~A.}\ \bibnamefont
  {Brun}},\ }\bibfield  {title} {\bibinfo {title} {Weak measurements are
  universal},\ }\href@noop {} {\bibfield  {journal} {\bibinfo  {journal}
  {Physical review letters}\ }\textbf {\bibinfo {volume} {95}},\ \bibinfo
  {pages} {110409} (\bibinfo {year} {2005})}\BibitemShut {NoStop}%
\bibitem [{\citenamefont {Tamir}\ and\ \citenamefont
  {Cohen}(2013)}]{tamir2013introduction}%
  \BibitemOpen
  \bibfield  {author} {\bibinfo {author} {\bibfnamefont {B.}~\bibnamefont
  {Tamir}}\ and\ \bibinfo {author} {\bibfnamefont {E.}~\bibnamefont {Cohen}},\
  }\bibfield  {title} {\bibinfo {title} {Introduction to weak measurements and
  weak values},\ }\href@noop {} {\bibfield  {journal} {\bibinfo  {journal}
  {Quanta}\ }\textbf {\bibinfo {volume} {2}},\ \bibinfo {pages} {7} (\bibinfo
  {year} {2013})}\BibitemShut {NoStop}%
\bibitem [{\citenamefont {Wu}\ \emph {et~al.}(2018)\citenamefont {Wu},
  \citenamefont {Coquet}, \citenamefont {Wang},\ and\ \citenamefont
  {Genevet}}]{wu2018modelling}%
  \BibitemOpen
  \bibfield  {author} {\bibinfo {author} {\bibfnamefont {K.}~\bibnamefont
  {Wu}}, \bibinfo {author} {\bibfnamefont {P.}~\bibnamefont {Coquet}}, \bibinfo
  {author} {\bibfnamefont {Q.~J.}\ \bibnamefont {Wang}},\ and\ \bibinfo
  {author} {\bibfnamefont {P.}~\bibnamefont {Genevet}},\ }\bibfield  {title}
  {\bibinfo {title} {Modelling of free-form conformal metasurfaces},\
  }\href@noop {} {\bibfield  {journal} {\bibinfo  {journal} {Nature
  communications}\ }\textbf {\bibinfo {volume} {9}},\ \bibinfo {pages} {3494}
  (\bibinfo {year} {2018})}\BibitemShut {NoStop}%
\bibitem [{\citenamefont {Lundeen}\ \emph {et~al.}(2009)\citenamefont
  {Lundeen}, \citenamefont {Feito}, \citenamefont {Coldenstrodt-Ronge},
  \citenamefont {Pregnell}, \citenamefont {Silberhorn}, \citenamefont {Ralph},
  \citenamefont {Eisert}, \citenamefont {Plenio},\ and\ \citenamefont
  {Walmsley}}]{lundeen2009tomography}%
  \BibitemOpen
  \bibfield  {author} {\bibinfo {author} {\bibfnamefont {J.~S.}\ \bibnamefont
  {Lundeen}}, \bibinfo {author} {\bibfnamefont {A.}~\bibnamefont {Feito}},
  \bibinfo {author} {\bibfnamefont {H.}~\bibnamefont {Coldenstrodt-Ronge}},
  \bibinfo {author} {\bibfnamefont {K.~L.}\ \bibnamefont {Pregnell}}, \bibinfo
  {author} {\bibfnamefont {C.}~\bibnamefont {Silberhorn}}, \bibinfo {author}
  {\bibfnamefont {T.~C.}\ \bibnamefont {Ralph}}, \bibinfo {author}
  {\bibfnamefont {J.}~\bibnamefont {Eisert}}, \bibinfo {author} {\bibfnamefont
  {M.~B.}\ \bibnamefont {Plenio}},\ and\ \bibinfo {author} {\bibfnamefont
  {I.~A.}\ \bibnamefont {Walmsley}},\ }\bibfield  {title} {\bibinfo {title}
  {Tomography of quantum detectors},\ }\href@noop {} {\bibfield  {journal}
  {\bibinfo  {journal} {Nature Physics}\ }\textbf {\bibinfo {volume} {5}},\
  \bibinfo {pages} {27} (\bibinfo {year} {2009})}\BibitemShut {NoStop}%
\bibitem [{\citenamefont {Ren}\ and\ \citenamefont
  {Li}(2019)}]{PhysRevLett.123.140405}%
  \BibitemOpen
  \bibfield  {author} {\bibinfo {author} {\bibfnamefont {H.}~\bibnamefont
  {Ren}}\ and\ \bibinfo {author} {\bibfnamefont {Y.}~\bibnamefont {Li}},\
  }\bibfield  {title} {\bibinfo {title} {Modeling quantum devices and the
  reconstruction of physics in practical systems},\ }\href
  {https://doi.org/10.1103/PhysRevLett.123.140405} {\bibfield  {journal}
  {\bibinfo  {journal} {Phys. Rev. Lett.}\ }\textbf {\bibinfo {volume} {123}},\
  \bibinfo {pages} {140405} (\bibinfo {year} {2019})}\BibitemShut {NoStop}%
\bibitem [{\citenamefont {Shaffer}\ \emph {et~al.}(2023)\citenamefont
  {Shaffer}, \citenamefont {Ren}, \citenamefont {Dyrenkova}, \citenamefont
  {Yale}, \citenamefont {Lobser}, \citenamefont {Burch}, \citenamefont {Chow},
  \citenamefont {Revelle}, \citenamefont {Clark},\ and\ \citenamefont
  {H{\"{a}}ffner}}]{Shaffer2023sampleefficient}%
  \BibitemOpen
  \bibfield  {author} {\bibinfo {author} {\bibfnamefont {R.}~\bibnamefont
  {Shaffer}}, \bibinfo {author} {\bibfnamefont {H.}~\bibnamefont {Ren}},
  \bibinfo {author} {\bibfnamefont {E.}~\bibnamefont {Dyrenkova}}, \bibinfo
  {author} {\bibfnamefont {C.~G.}\ \bibnamefont {Yale}}, \bibinfo {author}
  {\bibfnamefont {D.~S.}\ \bibnamefont {Lobser}}, \bibinfo {author}
  {\bibfnamefont {A.~D.}\ \bibnamefont {Burch}}, \bibinfo {author}
  {\bibfnamefont {M.~N.~H.}\ \bibnamefont {Chow}}, \bibinfo {author}
  {\bibfnamefont {M.~C.}\ \bibnamefont {Revelle}}, \bibinfo {author}
  {\bibfnamefont {S.~M.}\ \bibnamefont {Clark}},\ and\ \bibinfo {author}
  {\bibfnamefont {H.}~\bibnamefont {H{\"{a}}ffner}},\ }\bibfield  {title}
  {\bibinfo {title} {Sample-efficient verification of
  continuously-parameterized quantum gates for small quantum processors},\
  }\href {https://doi.org/10.22331/q-2023-05-04-997} {\bibfield  {journal}
  {\bibinfo  {journal} {{Quantum}}\ }\textbf {\bibinfo {volume} {7}},\ \bibinfo
  {pages} {997} (\bibinfo {year} {2023})}\BibitemShut {NoStop}%
\bibitem [{\citenamefont {Multiphysics}(1998)}]{multiphysics1998introduction}%
  \BibitemOpen
  \bibfield  {author} {\bibinfo {author} {\bibfnamefont {C.}~\bibnamefont
  {Multiphysics}},\ }\bibfield  {title} {\bibinfo {title} {Introduction to
  comsol multiphysics{\textregistered}},\ }\href@noop {} {\bibfield  {journal}
  {\bibinfo  {journal} {COMSOL Multiphysics, Burlington, MA, accessed Feb}\
  }\textbf {\bibinfo {volume} {9}},\ \bibinfo {pages} {2018} (\bibinfo {year}
  {1998})}\BibitemShut {NoStop}%
\bibitem [{\citenamefont {Bhattacharyya}(1946)}]{bhatt1946measure}%
  \BibitemOpen
  \bibfield  {author} {\bibinfo {author} {\bibfnamefont {A.}~\bibnamefont
  {Bhattacharyya}},\ }\bibfield  {title} {\bibinfo {title} {On a measure of
  divergence between two multinomial populations},\ }\href@noop {} {\bibfield
  {journal} {\bibinfo  {journal} {Sankhy{\=a}: the indian journal of
  statistics}\ }\textbf {\bibinfo {volume} {7}},\ \bibinfo {pages} {401}
  (\bibinfo {year} {1946})}\BibitemShut {NoStop}%
\bibitem [{\citenamefont {Chen}\ \emph {et~al.}(2021)\citenamefont {Chen},
  \citenamefont {Yu}, \citenamefont {Zeng},\ and\ \citenamefont
  {Flammia}}]{chen2021robust}%
  \BibitemOpen
  \bibfield  {author} {\bibinfo {author} {\bibfnamefont {S.}~\bibnamefont
  {Chen}}, \bibinfo {author} {\bibfnamefont {W.}~\bibnamefont {Yu}}, \bibinfo
  {author} {\bibfnamefont {P.}~\bibnamefont {Zeng}},\ and\ \bibinfo {author}
  {\bibfnamefont {S.~T.}\ \bibnamefont {Flammia}},\ }\bibfield  {title}
  {\bibinfo {title} {Robust shadow estimation},\ }\href@noop {} {\bibfield
  {journal} {\bibinfo  {journal} {PRX Quantum}\ }\textbf {\bibinfo {volume}
  {2}},\ \bibinfo {pages} {030348} (\bibinfo {year} {2021})}\BibitemShut
  {NoStop}%
\bibitem [{\citenamefont {Koh}\ and\ \citenamefont
  {Grewal}(2022)}]{koh2022classical}%
  \BibitemOpen
  \bibfield  {author} {\bibinfo {author} {\bibfnamefont {D.~E.}\ \bibnamefont
  {Koh}}\ and\ \bibinfo {author} {\bibfnamefont {S.}~\bibnamefont {Grewal}},\
  }\bibfield  {title} {\bibinfo {title} {Classical shadows with noise},\
  }\href@noop {} {\bibfield  {journal} {\bibinfo  {journal} {Quantum}\ }\textbf
  {\bibinfo {volume} {6}},\ \bibinfo {pages} {776} (\bibinfo {year}
  {2022})}\BibitemShut {NoStop}%
\bibitem [{\citenamefont {D{\"u}r}\ \emph {et~al.}(2000)\citenamefont
  {D{\"u}r}, \citenamefont {Vidal},\ and\ \citenamefont
  {Cirac}}]{dur2000three}%
  \BibitemOpen
  \bibfield  {author} {\bibinfo {author} {\bibfnamefont {W.}~\bibnamefont
  {D{\"u}r}}, \bibinfo {author} {\bibfnamefont {G.}~\bibnamefont {Vidal}},\
  and\ \bibinfo {author} {\bibfnamefont {J.~I.}\ \bibnamefont {Cirac}},\
  }\bibfield  {title} {\bibinfo {title} {Three qubits can be entangled in two
  inequivalent ways},\ }\href@noop {} {\bibfield  {journal} {\bibinfo
  {journal} {Physical Review A}\ }\textbf {\bibinfo {volume} {62}},\ \bibinfo
  {pages} {062314} (\bibinfo {year} {2000})}\BibitemShut {NoStop}%
\bibitem [{\citenamefont {Pradhan}\ \emph {et~al.}(2008)\citenamefont
  {Pradhan}, \citenamefont {Agrawal},\ and\ \citenamefont
  {Pati}}]{pradhan2008generalized}%
  \BibitemOpen
  \bibfield  {author} {\bibinfo {author} {\bibfnamefont {B.}~\bibnamefont
  {Pradhan}}, \bibinfo {author} {\bibfnamefont {P.}~\bibnamefont {Agrawal}},\
  and\ \bibinfo {author} {\bibfnamefont {A.}~\bibnamefont {Pati}},\ }\bibfield
  {title} {\bibinfo {title} {Generalized w-states and quantum communication
  protocols},\ }\href@noop {} {\bibfield  {journal} {\bibinfo  {journal} {arXiv
  preprint arXiv:0805.2651}\ } (\bibinfo {year} {2008})}\BibitemShut {NoStop}%
\bibitem [{\citenamefont {Meng}\ \emph {et~al.}(2023)\citenamefont {Meng},
  \citenamefont {Han}, \citenamefont {Cong},\ and\ \citenamefont
  {Guo}}]{MENG2023106661}%
  \BibitemOpen
  \bibfield  {author} {\bibinfo {author} {\bibfnamefont {X.}~\bibnamefont
  {Meng}}, \bibinfo {author} {\bibfnamefont {Z.}~\bibnamefont {Han}}, \bibinfo
  {author} {\bibfnamefont {J.}~\bibnamefont {Cong}},\ and\ \bibinfo {author}
  {\bibfnamefont {X.}~\bibnamefont {Guo}},\ }\bibfield  {title} {\bibinfo
  {title} {Intelligent optimization based density matrix reconstruction method
  with semi-positive constraint},\ }\href
  {https://doi.org/https://doi.org/10.1016/j.rinp.2023.106661} {\bibfield
  {journal} {\bibinfo  {journal} {Results in Physics}\ }\textbf {\bibinfo
  {volume} {51}},\ \bibinfo {pages} {106661} (\bibinfo {year}
  {2023})}\BibitemShut {NoStop}%
\bibitem [{\citenamefont {Cheng}\ \emph {et~al.}(2023)\citenamefont {Cheng},
  \citenamefont {Liang}, \citenamefont {Yang}, \citenamefont {Ren},
  \citenamefont {Shi}, \citenamefont {Li},\ and\ \citenamefont
  {Qian}}]{cheng2023fidelity}%
  \BibitemOpen
  \bibfield  {author} {\bibinfo {author} {\bibfnamefont {J.}~\bibnamefont
  {Cheng}}, \bibinfo {author} {\bibfnamefont {Z.}~\bibnamefont {Liang}},
  \bibinfo {author} {\bibfnamefont {R.}~\bibnamefont {Yang}}, \bibinfo {author}
  {\bibfnamefont {H.}~\bibnamefont {Ren}}, \bibinfo {author} {\bibfnamefont
  {Y.}~\bibnamefont {Shi}}, \bibinfo {author} {\bibfnamefont {T.}~\bibnamefont
  {Li}},\ and\ \bibinfo {author} {\bibfnamefont {X.}~\bibnamefont {Qian}},\
  }\href@noop {} {\bibinfo {title} {Fidelity estimator, randomized benchmarking
  and zne for quantum pulses}} (\bibinfo {year} {2023}),\ \Eprint
  {https://arxiv.org/abs/2305.12597} {arXiv:2305.12597 [quant-ph]} \BibitemShut
  {NoStop}%
\bibitem [{\citenamefont {Bose}\ \emph {et~al.}(2001)\citenamefont {Bose},
  \citenamefont {Fuentes-Guridi}, \citenamefont {Knight},\ and\ \citenamefont
  {Vedral}}]{PhysRevLett.87.050401}%
  \BibitemOpen
  \bibfield  {author} {\bibinfo {author} {\bibfnamefont {S.}~\bibnamefont
  {Bose}}, \bibinfo {author} {\bibfnamefont {I.}~\bibnamefont
  {Fuentes-Guridi}}, \bibinfo {author} {\bibfnamefont {P.~L.}\ \bibnamefont
  {Knight}},\ and\ \bibinfo {author} {\bibfnamefont {V.}~\bibnamefont
  {Vedral}},\ }\bibfield  {title} {\bibinfo {title} {Subsystem purity as an
  enforcer of entanglement},\ }\href
  {https://doi.org/10.1103/PhysRevLett.87.050401} {\bibfield  {journal}
  {\bibinfo  {journal} {Phys. Rev. Lett.}\ }\textbf {\bibinfo {volume} {87}},\
  \bibinfo {pages} {050401} (\bibinfo {year} {2001})}\BibitemShut {NoStop}%
\bibitem [{\citenamefont {Hu}\ \emph {et~al.}(2024)\citenamefont {Hu},
  \citenamefont {Gu}, \citenamefont {Majumder}, \citenamefont {Ren},
  \citenamefont {Zhang}, \citenamefont {Wang}, \citenamefont {You},
  \citenamefont {Minev}, \citenamefont {Yelin},\ and\ \citenamefont
  {Seif}}]{hu2024demonstration}%
  \BibitemOpen
  \bibfield  {author} {\bibinfo {author} {\bibfnamefont {H.-Y.}\ \bibnamefont
  {Hu}}, \bibinfo {author} {\bibfnamefont {A.}~\bibnamefont {Gu}}, \bibinfo
  {author} {\bibfnamefont {S.}~\bibnamefont {Majumder}}, \bibinfo {author}
  {\bibfnamefont {H.}~\bibnamefont {Ren}}, \bibinfo {author} {\bibfnamefont
  {Y.}~\bibnamefont {Zhang}}, \bibinfo {author} {\bibfnamefont {D.~S.}\
  \bibnamefont {Wang}}, \bibinfo {author} {\bibfnamefont {Y.-Z.}\ \bibnamefont
  {You}}, \bibinfo {author} {\bibfnamefont {Z.}~\bibnamefont {Minev}}, \bibinfo
  {author} {\bibfnamefont {S.~F.}\ \bibnamefont {Yelin}},\ and\ \bibinfo
  {author} {\bibfnamefont {A.}~\bibnamefont {Seif}},\ }\href@noop {} {\bibinfo
  {title} {Demonstration of robust and efficient quantum property learning with
  shallow shadows}} (\bibinfo {year} {2024}),\ \Eprint
  {https://arxiv.org/abs/2402.17911} {arXiv:2402.17911 [quant-ph]} \BibitemShut
  {NoStop}%
\bibitem [{\citenamefont {Moharam}\ \emph {et~al.}(1995)\citenamefont
  {Moharam}, \citenamefont {Grann}, \citenamefont {Pommet},\ and\ \citenamefont
  {Gaylord}}]{moharam1995formulation}%
  \BibitemOpen
  \bibfield  {author} {\bibinfo {author} {\bibfnamefont {M.}~\bibnamefont
  {Moharam}}, \bibinfo {author} {\bibfnamefont {E.~B.}\ \bibnamefont {Grann}},
  \bibinfo {author} {\bibfnamefont {D.~A.}\ \bibnamefont {Pommet}},\ and\
  \bibinfo {author} {\bibfnamefont {T.}~\bibnamefont {Gaylord}},\ }\bibfield
  {title} {\bibinfo {title} {Formulation for stable and efficient
  implementation of the rigorous coupled-wave analysis of binary gratings},\
  }\href@noop {} {\bibfield  {journal} {\bibinfo  {journal} {JOSA a}\ }\textbf
  {\bibinfo {volume} {12}},\ \bibinfo {pages} {1068} (\bibinfo {year}
  {1995})}\BibitemShut {NoStop}%
\bibitem [{\citenamefont {Hugonin}\ and\ \citenamefont
  {Lalanne}(2021)}]{hugonin2021reticolo}%
  \BibitemOpen
  \bibfield  {author} {\bibinfo {author} {\bibfnamefont {J.~P.}\ \bibnamefont
  {Hugonin}}\ and\ \bibinfo {author} {\bibfnamefont {P.}~\bibnamefont
  {Lalanne}},\ }\bibfield  {title} {\bibinfo {title} {Reticolo software for
  grating analysis},\ }\href@noop {} {\bibfield  {journal} {\bibinfo  {journal}
  {arXiv preprint arXiv:2101.00901}\ } (\bibinfo {year} {2021})}\BibitemShut
  {NoStop}%
\bibitem [{\citenamefont {Scott}(2006)}]{scott2006tight}%
  \BibitemOpen
  \bibfield  {author} {\bibinfo {author} {\bibfnamefont {A.~J.}\ \bibnamefont
  {Scott}},\ }\bibfield  {title} {\bibinfo {title} {Tight informationally
  complete quantum measurements},\ }\href@noop {} {\bibfield  {journal}
  {\bibinfo  {journal} {Journal of Physics A: Mathematical and General}\
  }\textbf {\bibinfo {volume} {39}},\ \bibinfo {pages} {13507} (\bibinfo {year}
  {2006})}\BibitemShut {NoStop}%
\bibitem [{\citenamefont {Watrous}(2018)}]{watrous2018theory}%
  \BibitemOpen
  \bibfield  {author} {\bibinfo {author} {\bibfnamefont {J.}~\bibnamefont
  {Watrous}},\ }\href@noop {} {\emph {\bibinfo {title} {The theory of quantum
  information}}}\ (\bibinfo  {publisher} {Cambridge university press},\
  \bibinfo {year} {2018})\BibitemShut {NoStop}%
\bibitem [{\citenamefont {Gross}\ \emph {et~al.}(2015)\citenamefont {Gross},
  \citenamefont {Krahmer},\ and\ \citenamefont {Kueng}}]{gross2015partial}%
  \BibitemOpen
  \bibfield  {author} {\bibinfo {author} {\bibfnamefont {D.}~\bibnamefont
  {Gross}}, \bibinfo {author} {\bibfnamefont {F.}~\bibnamefont {Krahmer}},\
  and\ \bibinfo {author} {\bibfnamefont {R.}~\bibnamefont {Kueng}},\ }\bibfield
   {title} {\bibinfo {title} {A partial derandomization of phaselift using
  spherical designs},\ }\href@noop {} {\bibfield  {journal} {\bibinfo
  {journal} {Journal of Fourier Analysis and Applications}\ }\textbf {\bibinfo
  {volume} {21}},\ \bibinfo {pages} {229} (\bibinfo {year} {2015})}\BibitemShut
  {NoStop}%
\bibitem [{\citenamefont {Foreman}\ \emph {et~al.}(2015)\citenamefont
  {Foreman}, \citenamefont {Favaro},\ and\ \citenamefont
  {Aiello}}]{foreman2015optimal}%
  \BibitemOpen
  \bibfield  {author} {\bibinfo {author} {\bibfnamefont {M.~R.}\ \bibnamefont
  {Foreman}}, \bibinfo {author} {\bibfnamefont {A.}~\bibnamefont {Favaro}},\
  and\ \bibinfo {author} {\bibfnamefont {A.}~\bibnamefont {Aiello}},\
  }\bibfield  {title} {\bibinfo {title} {Optimal frames for polarization state
  reconstruction},\ }\href@noop {} {\bibfield  {journal} {\bibinfo  {journal}
  {Physical review letters}\ }\textbf {\bibinfo {volume} {115}},\ \bibinfo
  {pages} {263901} (\bibinfo {year} {2015})}\BibitemShut {NoStop}%
\end{thebibliography}%

\end{document}